\documentclass [aps,prL,amssymb,amsmath,twocolumn,showpacs]{revtex4-1}
\usepackage{microtype}
\usepackage{amsmath}
\usepackage{amssymb}
\usepackage{amsthm}
\usepackage{mathrsfs,mathbbol}
\usepackage{graphicx}
\usepackage{verbatim}
\usepackage{bm}
\usepackage[dvipsnames]{xcolor}
\usepackage{graphicx,epsfig,amsfonts,amssymb}

\usepackage[colorlinks,
linkcolor=blue,
anchorcolor=blue,
urlcolor=blue,
citecolor=blue
]{hyperref}

\usepackage{bm}
\usepackage{times}
\usepackage{lipsum}

\usepackage[all]{xy}

\newcommand{\be}{\begin{eqnarray}}
\newcommand{\ee}{\end{eqnarray}}
\newcommand{\la}{\langle}
\newcommand{\ra}{\rangle}

\begin{document}

\title{Dynamical characterization of topological phases beyond the minimal models}

\author{Xi Wu} 
\altaffiliation[]{These authors contributed equally to this work.}
\author {Panpan Fang}
\altaffiliation[]{These authors contributed equally to this work.}
\author{Fuxiang Li}
\email{fuxiangli@hnu.edu.cn}
\affiliation{School of Physics and Electronics, Hunan University, Changsha 410082, China}

\date{\today}
\begin{abstract}
	Dynamical characterization of topological phases under quantum quench dynamics has been demonstrated as a powerful and efficient tool. Previous studies have been focused on systems of which the Hamiltonian consists of matrices that commute with each other and satisfy Clifford algebra. In this work, we consider the characterization of topological phases with Hamiltonians that are beyond the minimal model. Specifically, the quantum quench dynamics of two types of layered systems is studied, of which the consisting matrices of Hamiltonians do not all satisfy Clifford algebra. We find that the terms which anti-commute with others can hold common band-inversion surfaces, which controls the topology of all the bands, but for other terms, there is no universal behavior and need to be treated case by case. 
\end{abstract}


\date{\today}

\maketitle

\section{Introduction} 
Topological phases of matter are originally defined in equilibrium in terms of topological invariants by the bulk Hamiltonian in momentum space  or by the number of edge states on the boundary of the material~\cite{hasan2010colloquium,qi2011topological}. These two approaches are related with each other by the so-called bulk-edge correspondence~\cite{hatsugai1993chern,Wen:2004ym}. 
In recent years, the bulk-edge correspondence is generalized into the so-called bulk-edge-hinge/corner correspondence, in which {the } states localized on the intersection of two boundaries are protected by the topology of the material. Such topological phases are called higher-order topological insulators~\cite{Benalcazar_2017,Schindler2018}. 
Meanwhile many studies show that topological structure of a system can also be probed by non-equilibrium quench dynamics both in theory~\cite{CaioMD2015,Ying2016,WangCe2017,HuHaiping2020,McGinleyMax2019,ChenXin2020} and experiment~\cite{Flaschner:2018aa,Bo2018Song,ShuaiChen2018,Song:2019aa,XinTao2020,NIU20211168}. Specifically, a different kind of bulk-edge correspondence, based on non-equilibrium dynamics, has been proposed by Liu {\it et al.} through the so-called dynamical spin-texture fields on the band-inversion surfaces (BIS) in the time-averaged spin polarization (TASP). It has been formulated in theory~\cite{ZHANG20181385,ZhangLong2019,YuXiangLong2021}, and realized in the experiment~\cite{YaWangXJLiu2019,YiChangRuiXJliu2019}. Later, it was generalized into Floquet systems~\cite{ZhangLong2020} and higher-order topological insulators~\cite{LI20211502}. Compared with traditional equilibrium approaches, Liu's method not only shows band topology {from} a different angle, but also has high feasibility and experimental accuracy.

However, most of the non-equilibrium approaches carried so far on the topological systems has focused on simple models like Su-Schrieffer-Heeger (SSH) model~\cite{Su:1979ut}, Haldane model (or Chern insulator)~\cite{Haldane:1988uf}. Liu's studies also only consider examples of minimal models in which the Hamiltonians can be expanded in terms of Gamma matrices, all satisfying Clifford algebra. In such minimal models, topological invariant of the band can be expressed by the coefficient of the Gamma matrices in a very simple way and therefore BISs and spin-texture fields come in a natural way. One way of extending the minimal model is in the higher-order topological insulators by the nested configurations of the BIS~\cite{LI20211502,LI2}. It was mentioned that models beyond the minimal model is possible in such models. However, one still wonders how far the procedure of minimal model can be applied in the case of generic multi-band Hamiltonians.

In this paper, we  generalize Liu's idea  and consider systems that are beyond Liu's minimal models. In these systems, the Hamiltonians are expanded by matrices that do not satisfy Clifford algebra:  some terms anti-commute with others but some do not. Specifically, we consider  layered systems with  the same type of layer but with two different types of stacking, giving rise to two different topological structures~\cite{MinHongki2008,MinHongki20082,RuiYu2010Sci,Hashimoto2016}. Though these models carry some similarities with the minimal models, they are more general. One of them (AB\&BA stacking) can be block-diagonalized and  therefore can be separated into subsystems, i.e., minimal models, but the other one (BA stacking) cannot.

{We describe the quench dynamics of the layered systems} and characterize the topology of these systems. Our findings are as following:
The terms which anti-commute with others can hold common BIS, and controls the topology of all the bands without {referring to the information of other terms}. { However, with the band dispersion considered, the condition for a term to have a BIS is relaxed into that if it is none zero at all momentum value, it can keep the gap open for whatever deformation of the other terms.} This holds for both AB\&BA stacking and BA stacking systems.
 Inside the subsystems of AB\&BA stacking, there exist BIS that control the topology of the subsystems. In order to clarify the whole topological number, these subsystems need to be treated separately.
 BA stacking systems need to be treated as a whole. In addition, the interlayer hopping can also be obtained from the TASP itself in these two stacking systems.
 
The paper is organized as follows: in Sec.~\ref{section:layered systems}, we introduce two stacking types of layered systems, AB\&BA and BA stacking and their topology; in Sec.~\ref{section:bilayer systems}, we show the way to quench the system and characterize topology for bilayer systems, providing the essence of the method; in Sec.~\ref{section:multi systems}, we generalize the results into multilayer systems, showing this method to be generically useful; in Sec.~\ref{section:summary}, we conclude the paper and discuss possible future directions.

\section{Layered systems}\label{section:layered systems}
In this section, we introduce the layered systems,  and discuss their band topology. Firstly, we consider the following Hamiltonian for the monolayer system:
\be
H_1=\left[\begin{array}{cc}
	h_3 & h_1-ih_2   \\h_1+ih_2  & -h_3   
\end{array}\right].\label{eq:monolayer}
\ee	
 This Hamiltonian can describe interactions within two spin-half states such as Qi-Wu-Zhang model~\cite{QWZ2006}, or sublattice degrees of freedom such as Haldane model~\cite{Haldane:1988uf}.
Then the two types of layered systems can be given. One is the AB\&BA stacking system, which means that the interlayer hopping has the direction both from the A of the first layer to the B of the second layer and the B of the first layer to the A of the second layer. The other is the BA stacking system, which means that the interlayer hopping is only from the B of the first layer to the A of the second layer (AB stacking system has the same spectrum). For convenience, we leave all the details of the derivation in the {Appendix}.

{\it The AB\&BA stacking system.} The bilayer AB\&BA stacking system is defined as:
\begin{small}
\be
\begin{split}
H^2_{AB\&BA}&=
\left[\begin{array}{cccc}
	h_3 & h_1-ih_2 & &t  \\h_1+ih_2  & -h_3 & t &  \\ & t & h_3 & h_1-ih_2 \\ t&  & h_1+ih_2 & -h_3
\end{array}\right]\\
&=\sum_{i=1}^3 h_i \mathbb{1} \otimes \sigma_i+t\sigma_1\otimes\sigma_1\,.
\label{H2ABBA}
\end{split}
\ee
\end{small}
$t$ is the interlayer hopping amplitude. 
\begin{figure}[htbp]
\centering
\epsfig{file=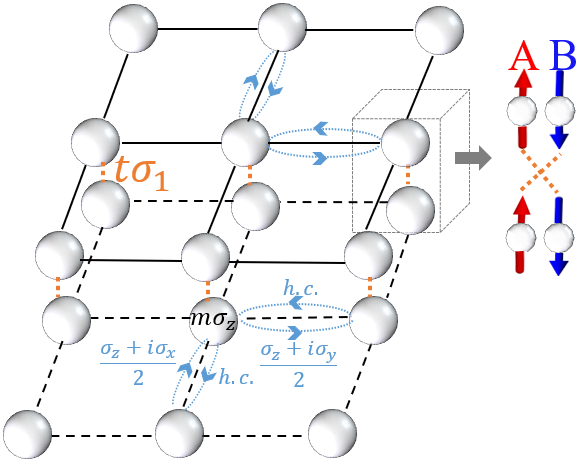, width=3.4in}
\caption{A lattice realization of $H^2_{AB\&BA}$ on a square bilayer lattice, in which up/down spin represent A/B degrees of freedom. The monolayer Hamiltonian is Qi-Wu-Zhang model, in which $h_1=\sin k_x$, $h_2=\sin k_y$, $h_3=m-\cos k_x-\cos k_y$.}
\label{fig:ABBA}
\end{figure}
The schematic diagram of this bilayer system is shown in Fig.~\ref{fig:ABBA}. The energy spectrum is
\be \label{SpeABBA2}
E^{\pm}_I&=&\pm\sqrt{(h_1+t)^2+h_2^2+h_3^2}=\pm E_I,\nonumber
\\
E^{\pm}_{II}&=&\pm\sqrt{(h_1-t)^2+h_2^2+h_3^2}=\pm E_{II}.
\ee

\begin{figure}[htbp]
\centering
\epsfig{file=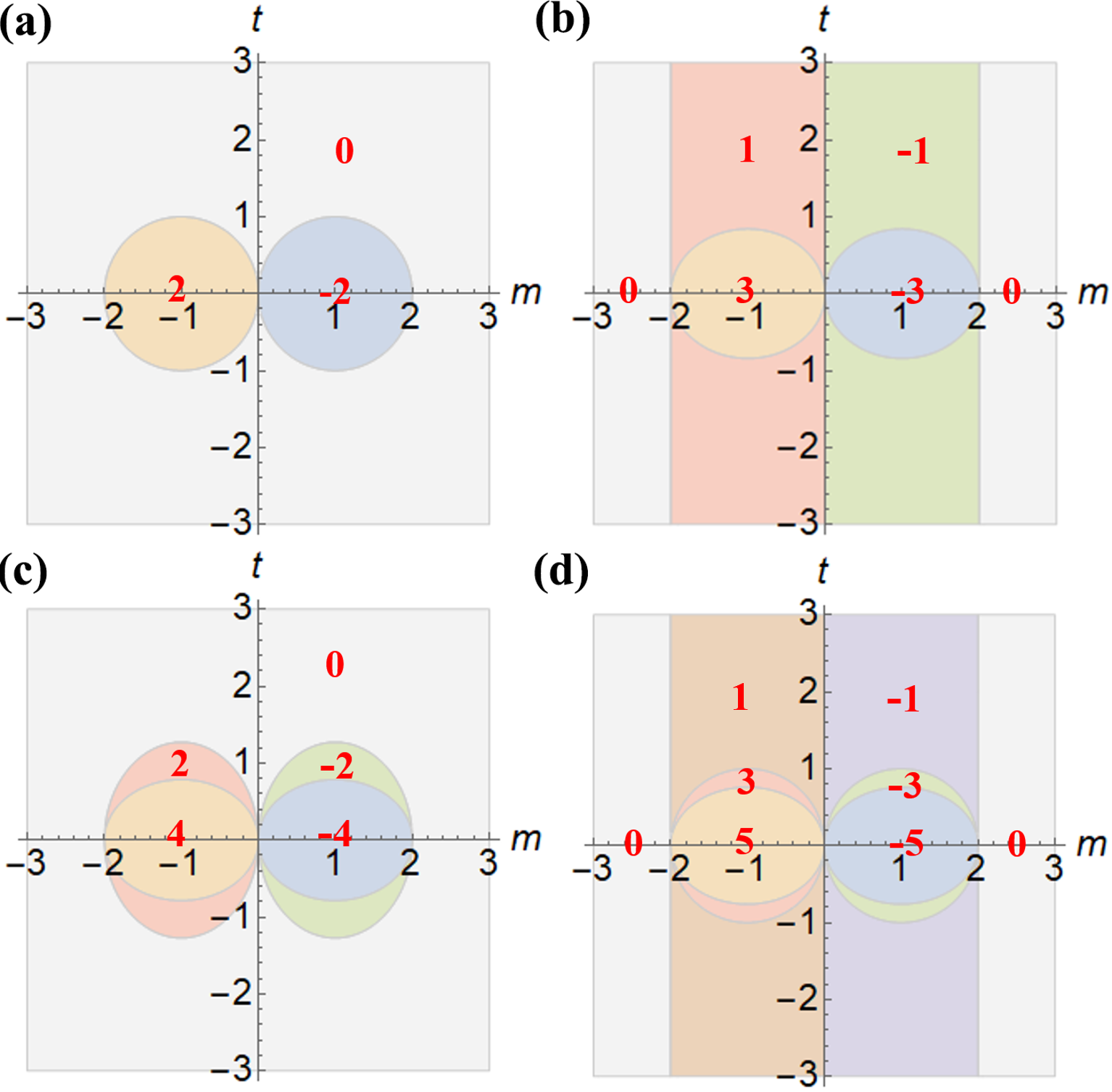, width=3.4in}
\caption{Phase diagrams of $H^N_{AB\&BA}$ with $h_1=\sin k_x$, $h_2=\sin k_y$, $h_3=m-\cos k_x-\cos k_y$. The {number of} layers $N$ in (a-d) corresponds to $2$-$5$, respectively. The numbers marked in red denote the topological number in corresponsing regions.}
\label{fig:pd1}
\end{figure}
{The Chern number of this bilayer system is the sum of the Chern numbers of the two subsystems $I$ and $II$. For details, see {\color{red}Appendix.~\ref{lam}}.} As shown in Fig.~\ref{fig:pd1}(a), we plot the phase diagram of above bilayer system.  One can see from the phase diagram: (i) If the monolayer system $H_1$ is topologically trivial, AB\&BA stacking system remains trivial for all $t$. (ii) If the monolayer model $H_1$ is topologically nontrivial, AB\&BA stacking system will undergo a phase transition by tuning $t$ at a finite $m$. { The phase transitions happen at $(t,m)$ such that $h_1\pm t=h_2=h_3=0$. } Thus, in order to focus on the effect of interlayer hopping on the layered systems,  $H_1$ is always assumed to be topological in the following.

The AB\&BA stacking multilayer models has the Hamiltonian:
\be
H^N_{AB\&BA}=\left[\begin{array}{cccc}\sum_{i=1}^3 h_i \sigma_i & t\sigma_1 &  &  \\t\sigma_1 & \sum_{i=1}^3 h_i & t\sigma_1 &  \\ & t\sigma_1 & \sum_{i=1}^3 h_i & ... \\ &  &...  & ...\end{array}\right]\label{HNABBA}
\ee
and its energy spectrum  is
\be  \label{SpeABBAn}
E^{\pm}_r=\pm\sqrt{(h_1-2t\cos\theta_r)^2+h_2^2+h_3^2}=\pm E_r.
\ee
with $\theta_r=\frac{r\pi}{N+1},r=1,2,...N$. {The Chern number of the multilayer system is the sum of all the Chern numbers of the subsystems labeled by $r$. For details, see {\color{red}Appendix.~\ref{lam}.} The multilayer system may also have phase transitions by tuning parameter $t$.  { The phase transitions happen at $(t,m)$ such that $h_1- 2t\cos\theta_r=h_2=h_3=0$ for each $r$. }If the monolayer system $H_1$ is topologically nontrivial, when tuning $t$ from $0$ to $\infty$, AB\&BA system will transit from a nontrivial topology into a final trivial one for even layered systems, and into {a topological phase with the absolute value of Chern number $1$} for odd layered systems; Otherwise, AB\&BA system remains trivial for all $t$. The phase diagram of the AB\&BA stacking multilayer system with $N=3,4$ and $5$ are given in Fig.~\ref{fig:pd1}(b), (c), and (d), respectively.

{\it The BA stacking system.} The bilayer BA stacking system is described by the Hamiltonian
\be
\begin{split}
H^2_{BA}&=
\left[\begin{array}{cccc}
	h_3 & h_1-ih_2 & &  \\h_1+ih_2  & -h_3 & t &  \\ & t & h_3 & h_1-ih_2 \\ &  & h_1+ih_2 & -h_3
\end{array}\right]\\
&
=\sum_{i=1}^3 h_i \mathbb{1} \otimes \sigma_i+\frac{t}{2}(\sigma_1\otimes\sigma_1+\sigma_2\otimes\sigma_2).
\label{H2BA}
\end{split}
\ee
\begin{figure}[htbp]
\centering
\epsfig{file=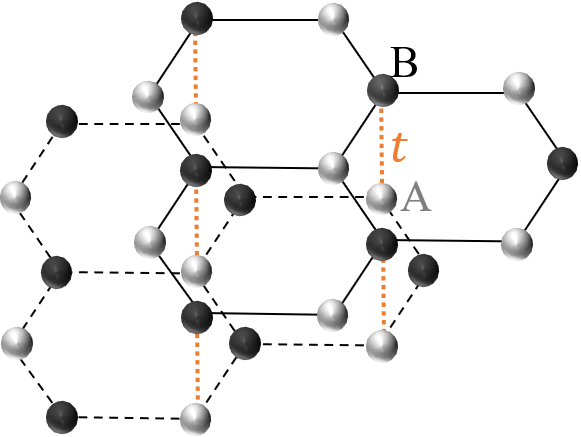, width=3.4in}
\caption{A lattice realization of $H^2_{BA}$ on a hexagonal bilayer lattice, in which white/black dots represent A/B sublattice degrees of freedom. The monolayer Hamiltonian describes Haldane model. }
\label{fig:BA}
\end{figure}
The schematic diagram of this bilayer system is shown in Fig.~\ref{fig:BA}. The energy spectrum is 
\be\label{SpeBA2}
E_1^{\pm}&=&\pm\sqrt{h_3^2+(\sqrt{(\frac{t}{2})^2+h_1^2+h_2^2}+\frac{t}{2})^2},\nonumber\\
E_2^{\pm}&=&\pm\sqrt{h_3^2+(\sqrt{(\frac{t}{2})^2+h_1^2+h_2^2}-\frac{t}{2})^2}.
\ee
\begin{figure}[htbp]
\centering
\epsfig{file=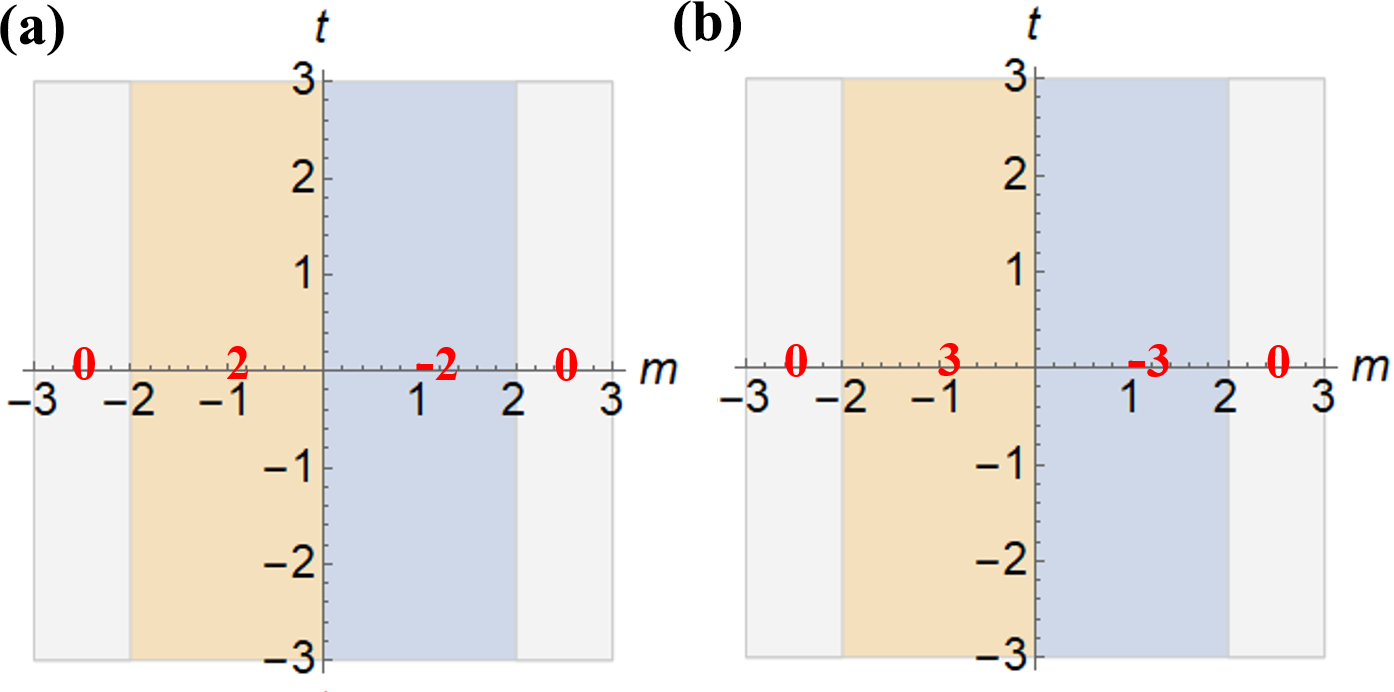, width=3.4in}
\caption{Phase diagrams of $H^2_{BA}$  and $H^3_{BA}$ with $h_1=\sin k_x$, $h_2=\sin k_y$, $h_3=m-\cos k_x-\cos k_y$. The numbers marked in red denote the topological number in corresponsing regions.}
\label{fig:pd2}
\end{figure}
in which  $E_1^+\geq E_2^+>E_2^-\geq E_1^-$,
and the ‘‘$=$" takes place at $t=0$. The spectrum tells us that at half-filling, the gap is not closed except at $t\to \infty$. Therefore, the Chern number of bilayer BA stacking model at any value of $t$ is the same as the one at $t=0$, which is the twice of the monolayer system. The phase diagram of above bilayer system is shown in Fig.~\ref{fig:pd2}(a). 

The multilayer BA stacking systems have many stacking types, here we only consider the one that is most common and stable in multilayer graphene, i.e., Bernal stacking. It is natural to believe that it is also stable in other hexagonal lattices. The multilayer BA stacking system has the $2N\times2N$ Hamiltonian $H^N_{BA}$
\begin{scriptsize}
\be
\left[\begin{array}{ccccccc}
	h_3 & h_1-ih_2 & &&&&  
	\\h_1+ih_2  & -h_3 & t & &&& 
	\\ & t & h_3 &  h_1-ih_2  && t&
	\\ &  & h_1+ih_2 & -h_3 &&&
	\\ &     &    &     &h_3 &  h_1-ih_2 & 
	\\ &     &   t &     &h_1+h_2&   -h_3&...
	\\ &     &            &  &&...& ...
\end{array}\right].\nonumber\\
\ee
\end{scriptsize}
The spectrum is 
\be\label{SpeBAn}
E_r^{\pm}&=&\pm\sqrt{h_3^2+(\sqrt{h_1^2+h_2^2+t^2\cos^2\theta_r}+t\cos\theta_r)^2}
\ee
with $\theta_r=\frac{r\pi}{N+1}$ and $r=1,2,...N$.
 The situation of multilayer BA stacking system is similar to the bilayer system at half-filling. The Chern number of BA stacking model at any value of $t$ is the same as the one at $t=0$, which is N times than the Chern number of the monolayer model. The phase diagram of the multilayer BA stacking model with $N=3$ is given in Fig.~\ref{fig:pd2}(b).
\section{Dynamical characterization in bilayer systems}\label{section:bilayer systems}
In this section, we show the way to characterize the bulk topology of the bilayer systems based on the TASP through quench dynamics. Unlike the minimal models, not all the terms in the Hamiltonian of layered systems satisfy Clifford algebra, and there is no standard mapping of the coefficients $h_i$ of these terms onto an effective field as in the two-level system. Therefore, the BIS, which emerges on the TASP, should be carefully defined. { For simplicity,  we use  the Hamiltonian of Qi-Wu-Zhang model for the Hamiltonian of monolayer model for the density plots in this section.}
\subsection{The common BIS}
\label{CMBIS}
In the case of layered systems, Clifford algebra is not satisfied for all terms in the Hamiltonian except a special term $h_3 \mathbb{1} \otimes \sigma_3$. The term $h_3 \mathbb{1} \otimes \sigma_3$ anti-commutes with all other terms and can be used to defined the common BIS in the AB\&BA stacking model as well as the bilayer BA stacking model. The processes are as follows:

Firstly, we prepare an initial state as:
\be
\rho_0&=&\frac{\mathbb{1}}{2}\otimes \frac{1}{2}(\mathbb{1}-\sigma_3)
=\frac{1}{2}\left[\begin{array}{cc}1 & 0 \\0 & 1\end{array}\right]\otimes\left[\begin{array}{cc}0 & 0 \\0 & 1\end{array}\right]\nonumber\\
&=&\frac{1}{2}\left[\begin{array}{cccc}0 & 0 & 0 & 0 \\0 & 1 & 0 & 0 \\0 & 0 & 0 & 0 \\0 & 0 & 0 & 1\end{array}\right],
\ee
which is a mixed state of two eigenstates of prequench Hamiltonian $H_0=h_3\mathbb{1} \otimes \sigma_3$ with the same eigenvalue. Then we can suddenly quench $H_0$ into postquench Hamiltonian $H_2$, which may be $H^2_{AB\&BA}$ or $H^2_{BA}$.  

After a long time of unitary evolution, one can obtain the corresponding TASP:
\be
\overline{\langle  \mathbb{1} \otimes \sigma_i \rangle}_{\rho_0}=-h_3 \sum_m \frac{\langle \widetilde{\psi}_m |  \mathbb{1} \otimes \sigma_i | \widetilde{\psi}_m\rangle}{4E_m},
\ee 
where $ | \widetilde{\psi}_m\rangle$ and $E_m$ are the normalized eigenvector and the eigenvalue of postquench Hamiltonian at $m$th-level, respectively. { \color{red}(See Appendix.~\ref{lem} for derivation.)} 

Thus, the BIS, where all the components of TASP vanish, can be identified at the region with momentum points:
\be
\text{BIS}=\{\mathbf{k}| \overline{\langle  \mathbb{1} \otimes \overrightarrow{ \sigma} \rangle}_{\rho_0}=0\}. 
\ee 
Or, from the perspective of Hamiltonian, the BIS is at the region with momentum points { where} $h_3=0$, which is consistent with the definition of BIS in Liu's paper~\cite{ZHANG20181385}. Especially, we have
\be
\overline{\langle  \mathbb{1} \otimes \sigma_3 \rangle}_{\rho_0}=-h_3^2 \sum_m \frac{1}{4E_m^2}.
\ee 
{\color{red}Let us explain the condition for BISs:  as far as there is some $h_i\ne0$ in the whole Brillouin zone which keeps the gap open regardless of other terms, we can continuously deform the Hamiltonian into $H=h_i \mathbb{1} \otimes \sigma_i$ which is a trivial one, so the disappearance of BIS means that the postquench Hamiltonian is topologically trivial. However, when there is a BIS appearing (or the surface where $h_i=0$) one is not allowed to deform Hamiltonian freely to $H=h_i \mathbb{1} \otimes \sigma_i$, therefore this may mean that the postquench Hamiltonian is topological. As one can see from Eq. (\ref{SpeABBA2}) (\ref{SpeABBAn}) (\ref{SpeBA2}) (\ref{SpeBAn}), $E^2$ is the sum of $h_3^2$ and other positive terms(See \cite{Wu2023TopHal} for general proof), so $h_3$ satisfies this condition quite obviously.} Of course, in order to know whether the system is truly topological we need to consider further information of the dynamical topological invariants, which is explained in the rest of this section. Thus, this BIS is as powerful as the one in the minimal models like monolayer system $H_1$ in (\ref{eq:monolayer}) for the characterization of bulk topology.  Here we need to emphasis that this special term $h_3 \mathbb{1} \otimes \sigma_3$ that anti-commutes with all other terms is a sufficient condition for having a BIS, but not a necessary one. As we will see in the following subsections that for other terms that do not anti-commute with all other terms, {\color{red} the characterization procedure depend very much on the energy spectrum considered. To be specific, for AB\&BA stacking model, $h_1\pm t$ and $h_2$ satisfy the condition for defining BIS while for BA stacking model $h_1$ and $h_2$ satisfy the condition above.}
 
\subsection{Dynamical characterization in AB\&BA stacking system}
In this subsection, we show how to capture bulk topology of the postquench Hamiltonian of the bilayer AB\&BA system. Starting from $\rho_0=\frac{\mathbb{1}}{2}\otimes \frac{1}{2}(\mathbb{1}-\sigma_j), j=1,2,3$, a mixed eigenstate of Hamiltonian $H_0=h_j\mathbb{1}\otimes\sigma_j$, we suddenly quench the Hamiltonian $H_0$ into ${H^2_{AB\&BA}}$ in (\ref{H2ABBA}). 
Then the exact form of TASP can be obtained as: 
\be
\overline{\langle  \mathbb{1} \otimes \sigma_i \rangle}_{\rho_0}=-\frac{h^i_Ih^j_I}{2E^2_I}-\frac{h^{i}_{II}h^{j}_{II}}{2E^2_{II}}, i=1,2,3.
\label{eq:twomixed}
\ee
in which $h^1_I=h_1+t$, $h^2_I=h_2$,  $h^3_I=h_3$, $h^1_{II}=h_1-t$, $h^2_{II}=h_2$ and $h^3_{II}=h_3$.

{However, Eq. (\ref{eq:twomixed}) gives the overlap between two topological structures, namely the BISs and the dynamical spin texture fields. BISs can be defined for $h_2$ and $h_3$ but not for $h_1$.} 
Therefore, in order to show the topological information independently, we consider splitting the TASP in  (\ref{eq:twomixed}) into two subspaces.
\begin{figure}[htbp]
	\centering
	\epsfig{file=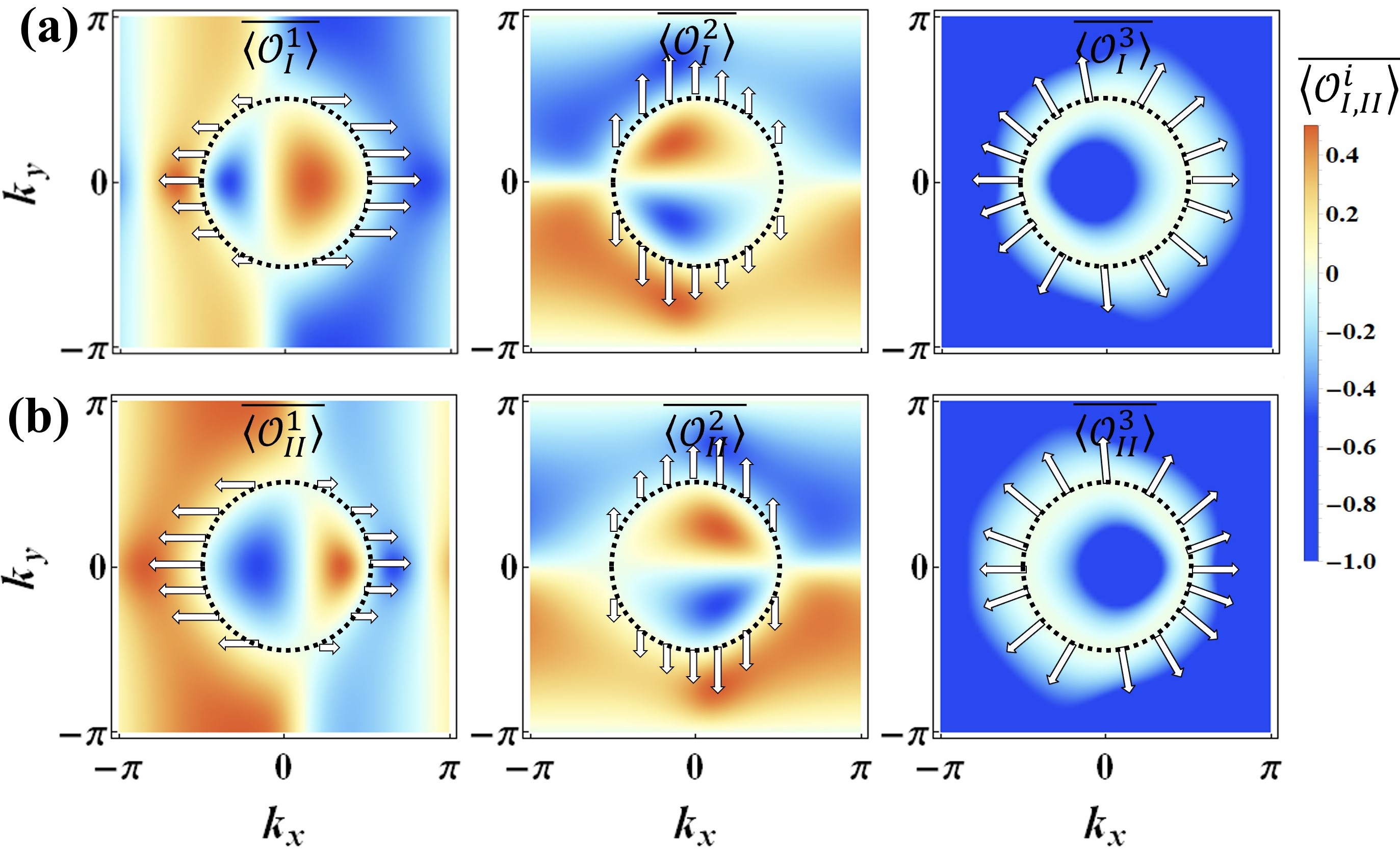, width=3.4in}
	\caption{(a) The GTASP and topological characterization of $H_I$. (b) The GTASP and topological characterization of $H_{II}$. Here, $m=1$, $t=0.4$ and $\rho_0=\frac{\mathbb{1}}{2}\otimes \frac{1}{2}(\mathbb{1}-\sigma_3)\,.$ The common rings emergent in (a) and (b) are the so-called BISs. { Along the BISs the winding of dynamical fields $\widetilde{\bm {g}_{I}}$ and $\widetilde{\bm {g}_{II}}$ give a nontrivial topological number $-1$. In the first two subfigures of (a) and (b) the white arrows represent the two components of $\widetilde{\bm {g}_{I}}$ and $\widetilde{\bm {g}_{II}}$ while the last subfigures represent  $\widetilde{\bm {g}_{I}}$ and $\widetilde{\bm {g}_{II}}$ as linear combinations of the components, respectively.}}
	\label{fig:bi1}
\end{figure} 

{By defining ${\mathcal D}_{\pm}=\frac{1}{2}(\mathbb{1}\pm\sigma_1)$, we obtain the long-time average of the operators  $\mathcal{O}^i_I=2{\mathcal D}_+\otimes \sigma_i$ and $\mathcal{O}^i_{II}=2{\mathcal D}_-\otimes \sigma_i$}, which we call GTASP (generalized time-averaged spin polarization) hereafter,  as follows:
\be
\overline{\langle {\mathcal{O}^i_I} \rangle}_{\rho_0}&=&-\frac{h^i_Ih^j_I}{E^2_I}.
\\
\overline{\langle  {\mathcal{O}^i_{II}} \rangle}_{\rho_0}&=&\frac{h^{i}_{II}h^{j}_{II}}{E^2_{II}}.
\label{eq:two}
\ee

At this time, the two BISs are well defined and identified in corresponding subspace:
\be
\text{BIS}_I=\{\mathbf{k}| \overline{\langle \overrightarrow{ \mathcal{O}_I} \rangle}_{\rho_0}=0\}.\\
\text{BIS}_{II}=\{\mathbf{k}| \overline{\langle \overrightarrow{ \mathcal{O}_{II}} \rangle}_{\rho_0}=0\}.
\ee 

Then after measuring the corresponding dynamical spin-texture fields, i.e., gradient fields of GTASP, on $\text{BIS}_I$ and $\text{BIS}_{II}$,  one can easily characterize the bulk topology of the system. The gradient fields in two subspaces are 
\be
\widetilde{g^{i}_I(\mathbf{k})}&=&
-\frac{1}{\mathcal{N}_\mathbf{k}}\partial_{k_{\bot}}\overline{\langle  {\mathcal{O}^{i}_I} \rangle}_{\rho_0}.
\\
\widetilde{g^{i}_{II}(\mathbf{k})}&=&
-\frac{1}{\mathcal{N}_\mathbf{k}}\partial_{k_{\bot}}\overline{\langle  {\mathcal{O}^{i}_{II}} \rangle}_{\rho_0}.
\ee
Here, $k_{\bot}$ is perpendicular to the BIS$_{I,II}$ 
and  {${\mathcal{N}_\mathbf{k}}$ is the normalization factor}.
After some algebras, we arrive at
\be
\widetilde{g^{i}_I(\mathbf{k})}\Big|_{\mathbf{k}\in  \text{BIS}_{I}}&=&\frac{h^i_I(0,\mathbf{k}_\parallel)}{\sum_{i\ne j} (h_I^i(0,\mathbf{k}_\parallel))^2}=\hat{h}^{so,i}_I.\\
\widetilde{g^{i}_{II}(\mathbf{k})}\Big|_{\mathbf{k}\in \text{BIS}_{II}}&=&\frac{h^i_{II}(0,\mathbf{k}_\parallel)}{\sum_{i\ne j} (h_{II}^i(0,\mathbf{k}_\parallel))^2}=\hat{h}^{so,i}_{II}.
\ee
Here, both $\widetilde{\bm {g}_{I}}$ and $\widetilde{\bm {g}_{II}}$ are the vectors with two components. If we choose the initial state $\rho_0=\frac{\mathbb{1}}{2}\otimes \frac{1}{2}(\mathbb{1}-\sigma_3)$,  ${\widetilde{\bm {g}_I}=\bm{\hat{h}^{so}}_I= (\hat h^1_I,\hat h^2_I)}$ and ${\widetilde{\bm {g}_{II}}=\bm{\hat{h}^{so}}_{II}= (\hat h^1_{II},\hat h^2_{II})}$.
 
The bulk topological number can be obtained by the sum of the dynamical invariants defined in each subspace along the BISs, 
\be
w=w_I+w_{II}=\frac{1}{2\pi}(\int_{\text{BIS}_I} \widetilde{\mathbf{g}_I} d \widetilde{\mathbf{g}_I}+\int_{\text{BIS}_{II}} \widetilde{\mathbf{g}_{II}} d \widetilde{\mathbf{g}_{II}}).\quad
\ee

As shown in Fig.~\ref{fig:bi1}(a) and (b), we plot the GTASP of $H_I$ and $H_{II}$ { for the case when the Chern number is $-2$ for the initial state $\rho_0=\frac{\mathbb{1}}{2}\otimes \frac{1}{2}(\mathbb{1}-\sigma_3)$ }, respectively. On the black dashed ring, all the components of GTASP vanish, and thus we identify it as BIS$_{I,II}$.  { The gradient fields $\widetilde{g_{I,II}(\mathbf{k})}$ (arrows) plotted along the BISs show topological number $-1$ for each subspace.}  
The bilayer system with a topological number $-2$ is successfully characterized. { In addition, using $h^1_I=h_1(\mathbf{k}) +t=0$, we can obtain $t$ as $-h^1(\mathbf{k}|_{h^1_I=0})$. This correspond to the  the white line across the BIS in the first plot of Fig.~\ref{fig:bi1}(a). Obviously, if the white line across the BIS disappears, meaning $h^1_I\ne0$ anywhere in the Brillouin zone, the model is topologically trivial. (See Appendix.~\ref{section:appendixa})}

\subsection{Dynamical characterization in BA stacking system}\label{section:BA_two}
\begin{figure}[htbp]
	\centering
	\epsfig{file=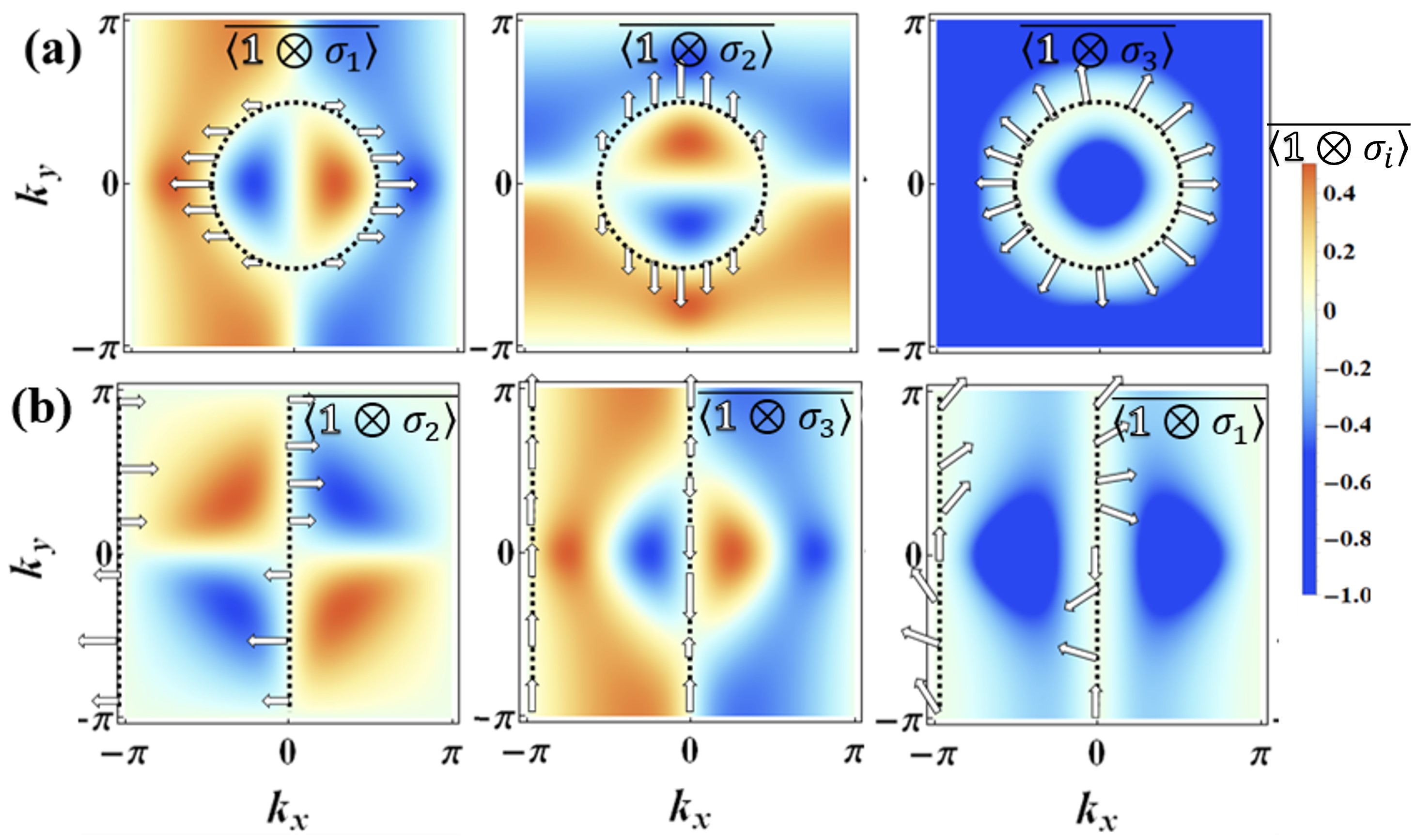, width=3.4in}
	\caption{(a-b) The TASP and topological characterization of $H^2_{BA}$.  Here, $m=1$ and $t=0.4$. The initial state $\rho_0$ in (a) and (b) are $\frac{\mathbb{1}}{2}\otimes \frac{1}{2}(\mathbb{1}-\sigma_3),$ $\frac{\mathbb{1}}{2}\otimes \frac{1}{2}(\mathbb{1}-\sigma_1)$, respectively.  { In the first two subfigures of (a) and (b) the white arrows represent the two components of $\widetilde{\bm g}$  while the last subfigures represent  $\widetilde{\bm g}$  as linear combinations of the components, respectively.}}
	\label{fig:bi2}
\end{figure}

 { The BA bilayer system is not a minimal model nor can it be  block diagonalized in a simple way}, meaning there is no independent subsystem as minimal models, so there is no simple way to calculate the Chern number directly. Similar to the previous subsection, for initial state $\rho_0=\frac{\mathbb{1}}{2}\otimes \frac{1}{2}(\mathbb{1}-\sigma_j), j=1,2,3.$,  we can obtain the TASP of bilayer system ${H^2_{BA}}$ as follows:
\be
	\overline{\langle  \mathbb{1} \otimes \sigma_i \rangle}_{\rho_0}=
-h_ih_jA_{ij},
\ee
in which  $A_{ij}$ are coefficients and  there is no summation over $i$ and $j$.   { Here fortunately we find the zeros of TASP remain the same as the monolayer model. The reason is that $t$ does not induce the gap closing so the topological structure remains the same. Thus
the BIS can be defined following the original minimal model}:
\be
\text{BIS }=\{\mathbf{k}| \overline{\langle  \mathbb{1} \otimes \overrightarrow{\sigma} \rangle}_{\rho_0}=0\}.
\ee 
{ One may wonder that why BISs still work for terms like $h_{1} \mathbb{1} \otimes \sigma_{1}$ and $h_{2} \mathbb{1} \otimes \sigma_{2}$ that do not anti-commute with all other terms. The reason can be seen in the dispersions Eq.(\ref{SpeBA2}) (\ref{SpeBAn}): When we keep the condition that  for instance $h_{1}\ne0$ throughout the Brillouin zone, the gap is not closed as we continously deform the Hamiltonian into $H=h_{1}\mathbb{1} \otimes \sigma_1$, which is a trivial one, and this shows that the original Hamiltonian is topologically trivial. Moreover, this condition works for all the BISs in AB\&BA systems as well. Therefore this  condition is necessary for a BIS to be useful for the models being considered in this paper.}

Next one measure the dynamical spin-texture fields on the BIS, which can be obtained from the gradient of TASP
\be
\widetilde{g_{i}(\mathbf{k})}=-\frac{1}{\mathcal{N}_\mathbf{k}}\partial_{k_{\bot}}\overline{\langle  \mathbb{1} \otimes \sigma_{i} \rangle}_{\rho_0},%
\ee
in which the difference is calculated as
\be
\Delta \overline{\langle  \mathbb{1} \otimes \sigma_{i} \rangle}_{\rho_0}\Big|_{k_{\bot}\to0}
&\propto&-2(A_{ij}h_{i})\Big|_{(0,\mathbf{k}_{\parallel})}k_\bot.
\ee
Thus, 
\be
\widetilde{g_{p}(\mathbf{k})}\Big|_{\mathbf{k}\in \text{BIS}}&=&\frac{A_{pj}h_p}{(A_{pj}h_p)^2+(A_{qj}h_q)^2}
\ee
with $p\ne q\ne j$ and $p,q,j=1,2,3$.  

Specifically, if we choose the initial state $\rho_0=\frac{\mathbb{1}}{2}\otimes\frac{1}{2}(\mathbb{1}-\sigma_3)$, $\widetilde{g_{1}}$ and $\widetilde{g_{2}}$ are equal to $\frac{A_{13}h_1}{(A_{13}h_1)^2+(A_{23}h_2)^2}$ and $\frac{A_{23}h_2}{(A_{23}h_2)^2+(A_{13}h_1)^2}$, respectively. 
Although there is a rescaling in $\widetilde{g_{i}}$, it does not change bulk topology. The topological number should be given by the winding of dynamical field on the BIS
	\be
	w=\frac{2}{2\pi}(\int_\text{BIS} \widetilde{\mathbf{g}} d \widetilde{\mathbf{g}}).
	\ee	
The TASP of the BA stacking system does not reflect the exact value of the Chern number, but only reflect whether the system is topological or not. {But since the only missing information is related with the number of layers which is two, we just added it as a coefficient by hand in the definition of $w$.}
  \begin{figure}[htbp]
\centering
\epsfig{file=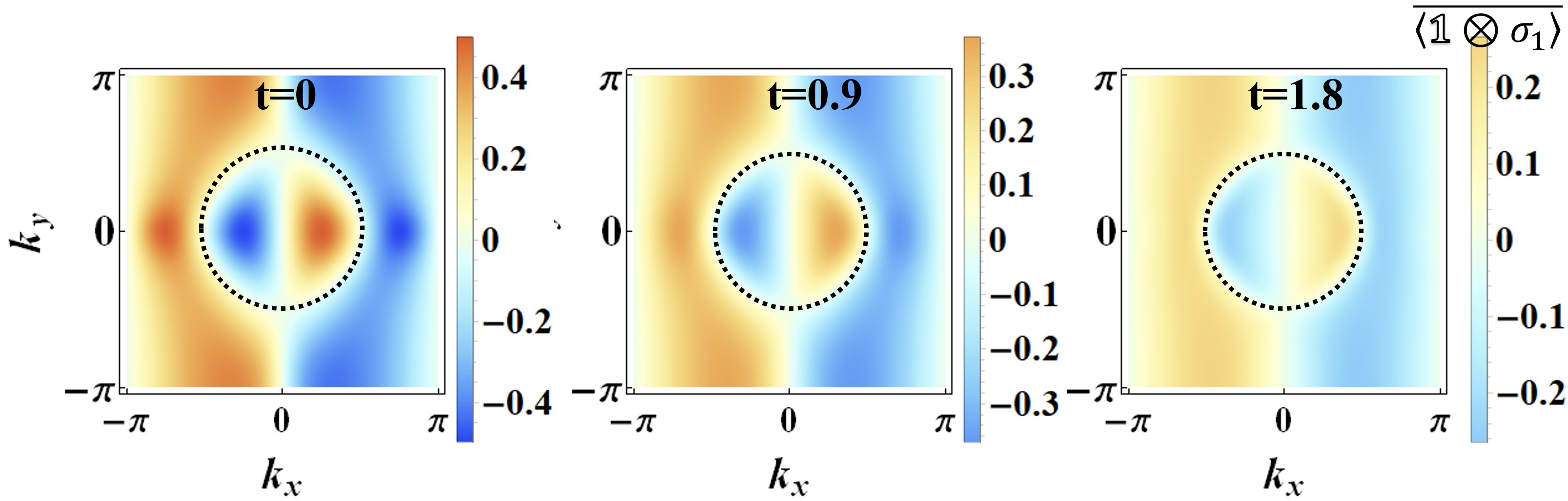, width=3.4in}
\caption{The component of TASP, $\overline{\langle  \mathbb{1} \otimes \sigma_1 \rangle}_{\rho_0}$ of $H^2_{BA}$ with different values of interlayer hopping $t$. }
\label{fig:quenchH2BAs1}
\end{figure}

{As shown in Fig.~\ref{fig:bi2}(a) and (b), we plot the TASP of  bilayer BA stacking system   for the case when the Chern number is $-2$ for $\rho_0=\frac{\mathbb{1}}{2}\otimes \frac{1}{2}(\mathbb{1}-\sigma_3)$ and $\rho_0=\frac{\mathbb{1}}{2}\otimes \frac{1}{2}(\mathbb{1}-\sigma_1)$, respectively. On dashed curves, all the components of TASP vanish, and thus we identify it as BIS. The gradient fields $\widetilde{g(\mathbf{k})}$ (arrows) plotted along the BISs show topological number $-2$.}

In addition, as shown in Fig.(\ref{fig:quenchH2BAs1}), take the initial state $\rho_0=\frac{\mathbb{1}}{2}\otimes \frac{1}{2}(\mathbb{1}-\sigma_3)$ as an example, we plot the component of TASP $\overline{\langle  \mathbb{1} \otimes \sigma_1 \rangle}_{\rho_0}$ with different interlayer hopping $t$. When increasing the interlayer hopping $t$, the maximum value of TASP decreases, and thus one can obtain the value of interlayer hopping $t$ by the variation of the amplitude of TASP  as shown in Fig.(\ref{fig:tmax}). 
If we choose the monolayer model as Haldane model, the results will be similar.(See Appendix.~\ref{section:appendixb})
\begin{figure}[htbp]
	\centering
	\epsfig{file=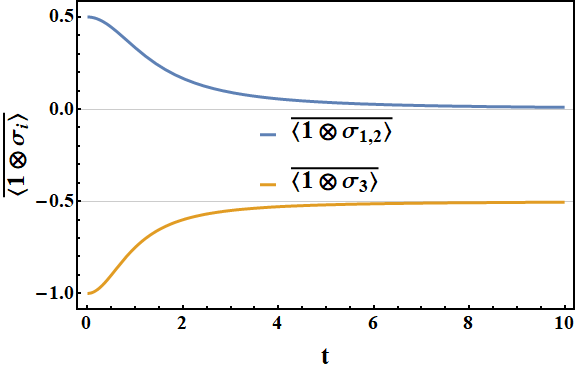, width=3.4in}
	\caption{The relation between the TASP $\overline{\langle  \mathbb{1} \otimes \bm \sigma \rangle}_{\rho_0}$  and interlayer hopping $t$. {We choose the momentum points, on which the correponding component of TASP have the maximum absolute value when $t=0$, to show the specific relation between the components of TASP and interlayer hopping $t$. Specifically, in $\overline{\langle  \mathbb{1} \otimes \sigma_1 \rangle}_{\rho_0}$, $\overline{\langle  \mathbb{1} \otimes \sigma_2 \rangle}_{\rho_0}$, and $\overline{\langle  \mathbb{1} \otimes \sigma_3 \rangle}_{\rho_0}$, we choose the momentum points $(\frac{\pi}{4},0)$, $(0,\frac{\pi}{4})$ and $(0,0)$, respectively. At the momentum points $(\frac{\pi}{4},0)$ and $(0,\frac{\pi}{4})$, both $\overline{\langle  \mathbb{1} \otimes \sigma_1 \rangle}_{\rho_0}$ and $\overline{\langle  \mathbb{1} \otimes \sigma_2 \rangle}_{\rho_0}$ are equal to $\frac{1}{2 + t^2}$. At the momentum point $(0,0)$, $\overline{\langle  \mathbb{1} \otimes \sigma_3 \rangle}_{\rho_0}$ is equal to $\frac{2 + t^2}{2 + 2t^2}$.}}
	\label{fig:tmax}
\end{figure}

\section{dynamical characterization in multilayer system}\label{section:multi systems}
In this section we are going to discuss the situation in the system with more layers than two. The common BIS is the same as in Sec. (\ref{CMBIS}) except that we need to replace $\rho_0=\frac{\mathbb{1}}{2}\otimes \frac{1}{2}(\mathbb{1}-\sigma_3)$ by $\rho_0=\frac{\mathbb{1}}{N}\otimes \frac{1}{2}(\mathbb{1}-\sigma_3)$,  so we are not going to repeat the discussion here. The other parts of the discussion in Sec.~\ref{section:bilayer systems} can be easily generalized in the following subsections. 
\subsection{Dynamical characterization in multilayer AB\&BA stacking system}
For N-layer AB\&BA stacking system, there is a way to quench the subsystems independently, too. The procedure is parallel to the bilayer case. Starting from $\rho_0=\frac{\mathbb{1}}{N}\otimes \frac{1}{2}(\mathbb{1}-\sigma_j)\,, j=1,2,3$, a mixed eigenstate of Hamiltonian $H_0=h_j\mathbb{1}\otimes\sigma_j$, we quench the Hamiltonian into ${H^N_{AB\&BA}}$ in Eq.(\ref{HNABBA}). 
The TASP are 
\be
\overline{\langle  \mathbb{1} \otimes \sigma_i \rangle}_{\rho_0}=-\sum_{r=1}^N\frac{h^i_rh^j_r}{2E^2_r}\,. \label{TAsinABBA}
\ee 
The result of TASP in Eq. (\ref{TAsinABBA}) gives the overlap of N topological structures. Instead we find the GTASP
\be
\overline{\langle  \mathbf{a}_r\mathbf{a}_r^T\otimes \sigma_i \rangle}_{\rho_0}&=&-\frac{h^i_rh^j_r}{E^2_r}\,,
\ee
in which $\mathbf{a}_r=\sqrt{\frac{2}{N+1}}\left(\begin{array}{cccc}\sin\theta_r,  \sin2\theta_r,  ..., \sin N\theta_r\end{array}\right)^T\label{ar}$ and $h^1_r=h_1-2t\cos\theta_r\,,
h^2_r=h_2\,,\,h^3_r=h_3\,$with $\theta_r=\frac{r\pi}{N+1}$ and $r=1,2,...N$. Each set ($ \mathbf{a}_r\mathbf{a}_r^T\otimes \sigma_1$, $ \mathbf{a}_r\mathbf{a}_r^T\otimes \sigma_2 $,$\mathbf{a}_r\mathbf{a}_r^T\otimes \sigma_3$) lives in a subspace that is orthogonal to others and the GTASP can characterize topology of each subspace. 

The BIS in each subspace is written as 
\be
\text{BIS}_r=\{\mathbf{k}| \overline{\langle \mathbf{a}_r\mathbf{a}_r^T\otimes \sigma_i  \rangle}_{\rho_0}=0\}.
\ee 
Or, the BIS$_{r}$ is at the region where $h^j_r=0$.

The dynamical spin-texture fields in each subspace can also be obtained by calculating
\be
\widetilde{g^{i}_r(\mathbf{k})}&=&
-\frac{1}{\mathcal{N}_\mathbf{k}}\partial_{k_{\bot}}\overline{\langle \mathbf{a}_r\mathbf{a}_r^T\otimes \sigma_i \rangle}_{\rho_0}.
\ee 
After some algebras, we get
\be
\widetilde{g^{i}_r(\mathbf{k})}\Big|_{\mathbf{k}\in \text{BIS}_r}&=&\frac{h^i_r(0,\mathbf{k}_\parallel)}{\sum_{i\ne j} (h_r^i(0,\mathbf{k}_\parallel))^2}=\hat{h}^{so,i}_r.
\ee
And the bulk topological number is the sum of the dynamical invariants defined in the subspaces as following
\be
w=\sum_{r=1}^N w_r=\sum_{r=1}^N \frac{1}{2\pi}\int_{BIS} \widetilde{\mathbf{g}_r} d \widetilde{\mathbf{g}_r}\,.
\ee 
\subsection{Dynamical characterization in multilayer BA stacking system}
Following the same procedure as in Sec.~\ref{section:BA_two}, starting from $\rho_0=\frac{\mathbb{1}}{N}\otimes \frac{1}{2}(\mathbb{1}-\sigma_j), j=1,2,3.$, we quench the Hamiltonian into ${H^N_{BA}}$, and then obtain the TASP. 
After some algebras, we have 
\be
\overline{\langle  \mathbb{1} \otimes \sigma_i \rangle}_{\rho_0}=
-h_ih_jA^{(N)}_{ij}
\ee
in which there is no summation over $i$ and $j$. $A^{(N)}_{ij}$ are the coefficients that depend on the numbers of layers. 

Like bilayer BA stacking system, the BIS can be defined as following:
\be
\text{BIS}=\{\mathbf{k}| \overline{\langle  \mathbb{1} \otimes \overrightarrow{\sigma} \rangle}_{\rho_0}=0\}.
\ee

Let us look at the dynamical spin-texture fields: 
\be
\widetilde{g_{i}(\mathbf{k})}=-\frac{1}{\mathcal{N}_\mathbf{k}}\partial_{k_{\bot}}\overline{\langle  \mathbb{1} \otimes \sigma_{i} \rangle}_{\rho_0}.
\ee
For initial state $$\rho_0=\frac{\mathbb{1}}{N}\otimes \frac{1}{2}(\mathbb{1}-\sigma_j),$$ whose BIS is at $h_j=0$, the difference is calculated as
\be
\Delta \overline{\langle  \mathbb{1} \otimes \sigma_{i} \rangle}_{\rho_0}\Big|_{k_{\bot}\to0}
&\propto&-2(A^{(N)}_{ij}h_{so,i})\Big|_{(0,\mathbf{k}_{\parallel})}k_\bot.
\ee
Thus, 
\be
\widetilde{g_{p}(\mathbf{k})}\Big|_{\mathbf{k}\in \text{BIS}}&=&\frac{A^{(N)}_{pj}h_p}{(A^{(N)}_{pj}h_p)^2+(A^{(N)}_{qj}h_q)^2},
\ee
with $p\ne q\ne j$ and $p,q,j=1,2,3$. 

The topological number of BA stacking N-layer system is N times than the monolayer case and there is no phase transition by tuning $t$ from zero to infinity. The TASP  do not reflect the value of topological number of multilayer BA stacking system because the structure of N layers overlaps with each other in the density plot in the Brillouin zone. But they do reflect whether the system is topological or not. The exact value of topological number for the N-layered system should be given by
\be
w=\frac{N}{2\pi}(\int_{BIS} \widetilde{\mathbf{g}} d \widetilde{\mathbf{g}}).
\ee

\section{Summary and discussions}\label{section:summary}
We studied the dynamical characterization of topology in two types of layered systems, which are beyond the minimal models. We found that the term that anti-commutes with all other terms still has the common BIS in this two layered systems. For AB\&BA system, because of block diagonalization, in order to fully describe the topology, we need to redefine BIS and new observables for spin-texture fields in the subspaces. While for BA system, the old BIS and observables are still working. {The condition for a term to have a BIS is relaxed into that if it is none zero at all momentum value, it can keep the gap open for whatever deformation of the other terms}. In addition, the magtitude of interlayer hopping can also be obtained from the TASP itself in these two stacking systems. 

Since we studied two types of models and they share similar  dynamical characterization of topology because of anti-commutation relation, it would be interesting to study whether we will have a general quench characterization of topology for a Hamiltonian that does not have a term that anti-commute with all the other terms. For instance, for a Hamiltonian expand by 3-by-3 Gell-Man matrices satisfying SU(3) algebra, the characterization will be more complicated. Thus, it is interesting to see how Liu's method can be applied. 

It was discussed in \cite{ZHANG20181385} that for models beyond the minimal models, it is possible to get a block-diagonalized form at each point in Brillouin zone,  and then BIS can be defined. However, our BA stacking model does not seem to fall into this category: from the dispersion Eq. (\ref{SpeBA2}) and (\ref{SpeBAn}), as far as $t$ is not zero, they are not like the spectrum of $\mathbf{\sigma} \cdot \mathbf{h}$ for some simple functions $\mathbf{h}$, and thus cannot be block diagonalizable in most regions of the Brillouin zone. 


\section*{Acknowledgements} 
This work was supported by the National Key Research and Development Program of the Ministry of Science and Technology (Grant No. 2021YFA1200700), the National Natural Science Foundation of China (Grant No. 12275075) and the Fundamental Research Funds for the Central Universities from China. 
\section{appendix}
\subsection{The GTASP of bilayer AB\&BA stacking system for the trivial postquench Hamiltonian}\label{section:appendixa}
As plotted in Fig.~\ref{fig:bi1_tr}, unlike the postquench Hamiltonian in main text, which lies in a topological regime, the components of GTASPs ${\overline{\la {\mathcal{O}^{1}_{I,II}}\ra}}$ for the trivial postquench Hamiltonian do not vanish on the line across the BIS. 
\begin{figure}[htbp]
	\centering
	\epsfig{file=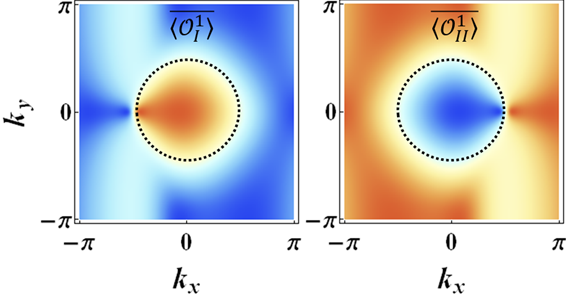, width=3.4in}
	\caption{The components of GTASPs ${\overline{\la {\mathcal{O}^{1}_{I,II}}\ra}}$. Here, $m=1$, $t=1.2$ and $\rho_0=\frac{\mathbb{1}}{2}\otimes \frac{1}{2}(\mathbb{1}-\sigma_3)$. }
	\label{fig:bi1_tr}
\end{figure} 
\subsection{The TASP of bilayer BA stacking system for the Haldane model}\label{section:appendixb}
As shown in Fig.~\ref{fig:ha}, we plot the TASP of bilayer BA stacking  system.  The monolayer Hamiltonian describes Haldane model, and its effective fields are $h_1=4\sum_{i=1}^3\cos (\bm {k\cdot a_i})$, $h_2=4\sum_{i=1}^3\sin (\bm {k\cdot a_i})$, and $h_3=m-2\sum_{i=1}^3\sin (\bm {k\cdot b_i})$. Here, $a_1=(0,1)$,  $a_2=(\frac{-\sqrt{3}}{2},\frac{-1}{2})$, $a_3=(\frac{\sqrt{3}}{2},\frac{-1}{2})$, $b_1=(-\sqrt{3},0)$, $b_2=(\frac{-\sqrt{3}}{2},\frac{3}{2})$, $b_3=(\frac{-\sqrt{3}}{2}\frac{-3}{2})$. The parameter $m=2\sqrt{3}$, which makes the postquench Hamiltonian lie in the topological regime. As we see, in each component of TASP, there exit three black dashed rings, on which all the components of TASP vanish. Thus, all the three rings can be identified as BIS. Taking one BIS (upper right) as an example, we measure the dynamical field (arrows) on this BIS in the components of TASP $\overline{\langle  \mathbb{1} \otimes \sigma_1 \rangle}_{\rho_0}$ and $\overline{\langle  \mathbb{1} \otimes \sigma_2 \rangle}_{\rho_0}$, respectively, and then combine them into the component $\overline{\langle  \mathbb{1} \otimes \sigma_3 \rangle}_{\rho_0}$. The winding of the dynamical field on this BIS manifests a nontrivial topological phase, which implies the system lies in the topological phase. If the postquench Hamiltonian lies in the trivial regime, there is no BIS appearing in the TASP.	
\begin{figure}[htbp]
	\centering
	\epsfig{file=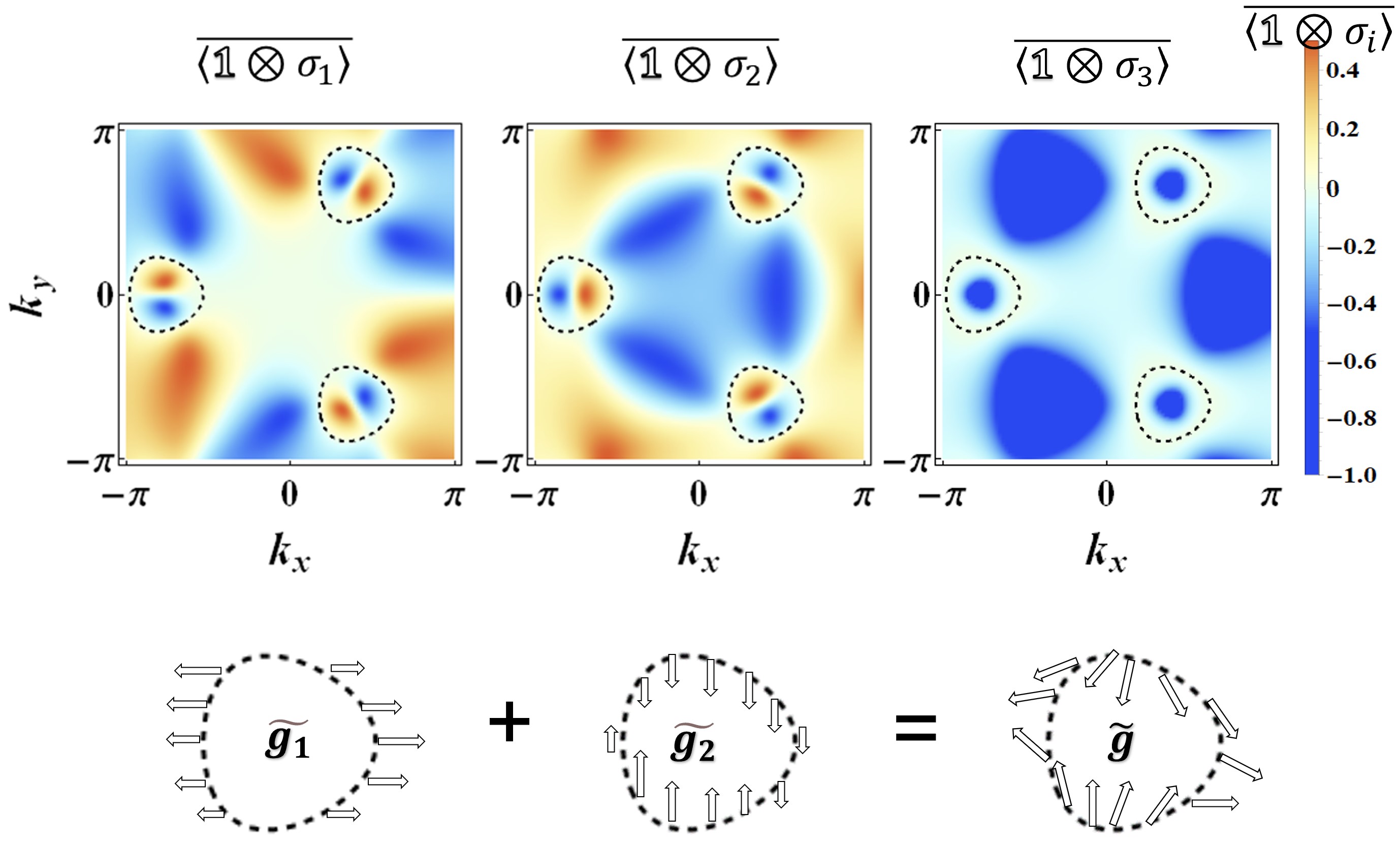, width=3.4in}
	\caption{The TASP and topological characterization of $H^2_{BA}$.  Here, the monolayer Hamiltonian describes Haldane model. The initial state $\rho_0$ is $\frac{\mathbb{1}}{2}\otimes \frac{1}{2}(\mathbb{1}-\sigma_3)$. To show the dynamical field better, the combined field $\widetilde g$ is normarnized. The interlayer hopping $t= 0.4$.}
	\label{fig:ha}
\end{figure} 
\subsection{Spectra of layered systems}\label{lam}
\subsubsection{AB\&BA stacking systems}
The AB\&BA stacking bilayer model is defined by the Hamiltonian
\be
H^2_{AB\&BA}&:=&
\left[\begin{array}{cccc}
	h_3 & h_1-ih_2 & &t  \\h_1+ih_2  & -h_3 & t &  \\ & t & h_3 & h_1-ih_2 \\ t&  & h_1+ih_2 & -h_3
\end{array}\right]\nonumber\\
&=&\sum_{i=1}^3 h_i \mathbb{1} \otimes \sigma_i+t\sigma_1\otimes\sigma_1,
\ee
in which the anti-commutation relations are 
\be\label{antiH2ABBA}
\{\mathbb{1}\otimes\sigma_i, \mathbb{1}\otimes\sigma_j\}&=&2\mathbb{1}\otimes\delta_{ij},\nonumber\\
\{\mathbb{1}\otimes\sigma_{2}, \sigma_1\otimes\sigma_1\}&=&0,\\
\{\mathbb{1}\otimes\sigma_{3}, \sigma_1\otimes\sigma_1\}&=&0.\nonumber
\ee
We block-diagonalize this Hamiltonian to get two subsystems:
\begin{small}
	\be\label{H2ABBA'}
	{H^2_{AB\&BA}}'&=&\frac{\mathbb{1}+i\sigma_2}{\sqrt{2}}\otimes \mathbb{1} H^2_{AB\&BA} \frac{\mathbb{1}-i\sigma_2}{\sqrt2}\otimes \mathbb{1}\nonumber\\
	&=&\frac{\mathbb{1}+i\sigma_2}{\sqrt{2}}\otimes \mathbb{1} (\sum_{i=1}^3 h_i \mathbb{1} \otimes \sigma_i+t\sigma_1\otimes\sigma_1) \frac{\mathbb{1}-i\sigma_2}{\sqrt2}\otimes \mathbb{1}\nonumber\\
	&=&\sum_{i=1}^3 h_i \mathbb{1} \otimes \sigma_i+t\sigma_3\otimes\sigma_1\nonumber\\
	&=& \left[\begin{array}{cccc}
		h_3 & h_1+t-ih_2 & &  \\h_1+t+ih_2  & -h_3 &  &  \\ &  & h_3 & h_1-t-ih_2 \\ &  & h_1-t+ih_2 & -h_3
	\end{array}\right]\nonumber\\
	&=:&H_I\oplus H_{II}\,.
	\ee
\end{small}
Therefore, the spectrum is
\be
E^{\pm}_I&=&\pm\sqrt{(h_1+t)^2+h_2^2+h_3^2}=\pm E_I,\nonumber
\\
E^{\pm}_{II}&=&\pm\sqrt{(h_1-t)^2+h_2^2+h_3^2}=\pm E_{II}.
\ee

The  multilayer AB\&BA stacking model has the Hamiltonian
\be
H^N_{AB\&BA}&=&\left[\begin{array}{cccc}\sum_{i=1}^3 h_i \sigma_i & t\sigma_1 &  &  \\t\sigma_1 & \sum_{i=1}^3 h_i & t\sigma_1 &  \\ & t\sigma_1 & \sum_{i=1}^3 h_i & ... \\ &  &...  & ...\end{array}\right]\nonumber\\
&=&
\sum_{i=1}^3 h_i \mathbb{1} \otimes \sigma_i+t\Sigma_1\otimes\sigma_1
\ee
where 
\be
\Sigma_1:=\left[\begin{array}{ccccc}
	0 & 1 & 0 & 0 &  
	\\1 & 0 & 1 & 0 &  
	\\0 & 1& 0 & 1 &  ...
	\\0 & 0 & 1 & 0 &  
	\\   &    & ... &    & 
\end{array}\right]\,.
\ee
and the anti-commutation relations are 
\be\label{antiNHABBA}
\{\mathbb{1}\otimes\sigma_i, \mathbb{1}\otimes\sigma_j\}&=&2\mathbb{1}\otimes\delta_{ij},\nonumber\\
\{\mathbb{1}\otimes\sigma_{2}, \Sigma_1\otimes\sigma_1\}&=&0,\\
\{\mathbb{1}\otimes\sigma_{3}, \Sigma_1\otimes\sigma_1\}&=&0.\nonumber
\ee
To block diagonalize $H^N_{AB\&BA}$, we just need to diagonalize $\Sigma_1$, which is given by
\be
\Sigma'_1=S_{\Sigma_1}^{-1}\Sigma_1 S_{\Sigma_1}=
\left[\begin{array}{cccc}
	2\cos\theta_1 &  &  &
	\\ & 2\cos\theta_2 &  &
	\\ &  & ...&
	\\ &  & & 2\cos\theta_N
\end{array}\right],\quad
\ee
where
\be\label{eq:ar}
S_{\Sigma_1}&=&\left(\begin{array}{cccc}\mathbf{a}_1, & \mathbf{a}_2, & ..., & \mathbf{a}_N\end{array}\right),\\\nonumber
\mathbf{a}_r&=&\sqrt{\frac{2}{N+1}}\left(\begin{array}{cccc}\sin\theta_r, & \sin2\theta_r, & ...,& \sin N\theta_r\end{array}\right)^T,\\\nonumber
\theta_r&=&\frac{r\pi}{N+1}, r=1,2,...N.
\ee
Therefore the block diagonalized Hamiltonian is
\be
{H^N_{AB\&BA}}'
=:\oplus_{r=1}^{n} H_r
\ee
where
\be
H_r=\sum_{i=1}^3 h_i \sigma_i-2t\cos\theta_r.
\ee	
and the spectrum is
\be
E^{\pm}_r=\pm\sqrt{(h_1-2t\cos\theta_r)^2+h_2^2+h_3^2}=\pm E_r, r=1,2,...N.
\ee
\subsubsection{BA stacking systems}
The bilayer BA stacking model is defined by the Hamiltonian
\be
&&H^2_{BA}\nonumber\\&=&
\left[\begin{array}{cccc}
	h_3 & h_1-ih_2 & &  \\h_1+ih_2  & -h_3 & t &  \\ & t & h_3 & h_1-ih_2 \\ &  & h_1+ih_2 & -h_3
\end{array}\right]\nonumber\\
&=&\sum_{i=1}^3 h_i \mathbb{1} \otimes \sigma_i+\frac{t}{2}(\sigma_1\otimes\sigma_1+\sigma_2\otimes\sigma_2).\nonumber\\
\ee
in which the anti-commutation relations are 
\be\label{antiH2AB}
\{\mathbb{1}\otimes\sigma_i, \mathbb{1}\otimes\sigma_j\}&=&2\mathbb{1}\otimes\delta_{ij},\nonumber\\
\{\mathbb{1}\otimes\sigma_3, \sigma_1\otimes\sigma_1\}&=&0,\\
\{\mathbb{1}\otimes\sigma_3,\sigma_2\otimes\sigma_2\}&=&0.\nonumber
\ee

The spectrum is 
\be
E_1^+&=&\sqrt{h_3^2+(\sqrt{(\frac{t}{2})^2+h_1^2+h_2^2}+\frac{t}{2})^2},\nonumber\\
E_2^+&=&\sqrt{h_3^2+(\sqrt{(\frac{t}{2})^2+h_1^2+h_2^2}-\frac{t}{2})^2},\nonumber\\
E_2^-&=&-\sqrt{h_3^2+(\sqrt{(\frac{t}{2})^2+h_1^2+h_2^2}-\frac{t}{2})^2},\\
E_1^-&=&-\sqrt{h_3^2+(\sqrt{(\frac{t}{2})^2+h_1^2+h_2^2}+\frac{t}{2})^2},\nonumber\\
E_1^+&\geq& E_2^+>E_2^-\geq E_1^-.\nonumber
\ee
All the eigenenergies satisfy 
\be
(E_m^2-\sum_ih_i^2)^2=t^2 (E_m^2-h_3^2)^2.
\ee
so the eigenvectors can be written as  
\be
|\psi_m\rangle=\left(\begin{array}{c}
	t(h_1-ih_2)(E_m+h_3) \\t(E_m^2-h_3^2) \\(E_m^2-\sum_ih_i^2)(E_m+h_3) \\(E_m^2-\sum_ih_i^2)(h_1+ih_2)
\end{array}\right),
\ee 
and the normalization is 
\be
\langle \psi_m |\psi_m\rangle&=&2t^2E_m(E_m+h_3)(E_m^2-h_3^2+h_1^2+h_2^2).\nonumber
\ee
Then we have
\be
\label{tpidsig1tp}\langle \widetilde{\psi}_m |  \mathbb{1} \otimes \sigma_1 | \widetilde{\psi}_m\rangle=\frac{2h_1(E_m^2-h_3^2)}{E_m(E_m^2-h_3^2+h_1^2+h_2^2)}.\\
\label{tpidsig2tp}\langle \widetilde{\psi}_m |  \mathbb{1} \otimes \sigma_2 | \widetilde{\psi}_m\rangle=\frac{2h_2(E_m^2-h_3^2)}{E_m(E_m^2-h_3^2+h_1^2+h_2^2)}.
\ee
and $\langle \widetilde{\psi}_m |  \mathbb{1} \otimes \sigma_3 | \widetilde{\psi}_m\rangle$is calculated in Eq. (\ref{idsig3}).

The multilayer BA stacking models have many types, here we only consider the one that is the most common and stable in multilayer graphene, Bernal stacking. It has the $2N\times2N$ Hamiltonian $H^N_{BA}$:
\begin{footnotesize}
	\be
	\left[\begin{array}{ccccccc}
		h_3 & h_1-ih_2 & &&&&  
		\\h_1+ih_2  & -h_3 & t & &&& 
		\\ & t & h_3 &  h_1-ih_2  && t&
		\\ &  & h_1+ih_2 & -h_3 &&&
		\\ &     &    &     &h_3 &  h_1-ih_2 & 
		\\ &     &   t &     &h_1+h_2&   -h_3&...
		\\ &     &            &  &&...& ...
	\end{array}\right],
	\ee
\end{footnotesize}
in which the anti-commutation relations are  
\be\label{antiHNAB}
\{\mathbb{1}\otimes\sigma_i, \mathbb{1}\otimes\sigma_j\}&=&2\mathbb{1}\otimes\delta_{ij},\nonumber\\
\{\mathbb{1}\otimes\sigma_3, \Sigma_1\otimes\sigma_1\}&=&0,\\
\{\mathbb{1}\otimes\sigma_3,\Sigma_2\otimes\sigma_2\}&=&0.\nonumber
\ee
where 
\be
\Sigma_2:=\left[\begin{array}{ccccc}
	0 & -i & 0 & 0 &  
	\\i & 0 & -i & 0 &  
	\\0 & i& 0 & -i &  ...
	\\0 & 0 & i & 0 &  
	\\   &    & ... &    & 
\end{array}\right].
\ee
The spectrum is 
\be
E_r^{\pm}&=&\pm\sqrt{h_3^2+(\sqrt{h_1^2+h_2^2+t^2\cos^2\theta_r}+t\cos\theta_r)^2},\nonumber\\
\theta_r&=&\frac{r\pi}{N+1}, r=1,2,...N.
\ee
All the energy eigenvalues satisfy 
\be
(E_r^2-\sum_ih_i^2)^2=4t^2 (E_r^2-h_3^2)^2 \cos^2\theta_r.
\ee
so the eigenket can be written as
\begin{footnotesize}  
	\be
	|\Psi_m\rangle^T&=&\Big((\psi_m^1)^T,(\psi_m^2)^T,...,(\psi_m^N)^T\Big),\nonumber
	\\ 
	|\psi_m^{2n-1}\rangle&=&\left(\begin{array}{c}
		2t(h_1-ih_2)(E_m+h_3)\cos\theta_r\sin((2n-1)\theta_r) \\2t(E_m^2-h_3^2)\cos\theta_r\sin((2n-1)\theta_r) \end{array}\right),\nonumber\\
	\\
	|\psi_m^{2n}\rangle&=&\left(\begin{array}{c}
		(E_m^2-\sum_ih_i^2)(E_m+h_3)\sin(2n\theta_r) \\(E_m^2-\sum_ih_i^2)(h_1+ih_2)\sin(2n\theta_r)
	\end{array}\right),\nonumber\\
	&=&\left(\begin{array}{c}
		2t(E_m+h_3)\sqrt{E_m^2-h_3^2}\cos\theta_r\sin(2n\theta_r) \\2t(h_1+ih_2)\sqrt{E_m^2-h_3^2}\cos\theta_r\sin(2n\theta_r)
	\end{array}\right).\nonumber
	\ee 
\end{footnotesize}
where $m$ labels both $\pm$ and $r$
and the normalization is 
\be
&&\langle \Psi_m |\Psi_m\rangle
\nonumber\\&=&2t^2\cos^2\theta_r(E_m+h_3)(E_m(E_m^2-h_3^2+h_1^2+h_2^2)(N+1)\nonumber
\\&&-h_3(E_m^2-h_3^2-h_1^2-h_2^2)(1+(-1)^{N+1})\,.
\ee
Then we have
\be
&&\langle \widetilde{\Psi}_m |  \mathbb{1} \otimes \sigma_1 | \widetilde{\Psi}_m\rangle
=\frac{2h_1(E_m^2-h_3^2)}{D_m}.\nonumber\\
&&\langle \widetilde{\Psi}_m |  \mathbb{1} \otimes \sigma_2 | \widetilde{\Psi}_m\rangle
=\frac{2h_2(E_m^2-h_3^2)}{D_m}.
\ee
where
\be
D_m&=&[E_m(E_m(E_m^2-h_3^2+h_1^2+h_2^2)\nonumber\\
&-&\frac{1+(-1)^{N+1}}{N+1}h_3(E_m^2-h_3^2-h_1^2-h_2^2))].
\ee
\subsection{Density matrices of the initial states}
\label{prequ}
We choose prequench Hamiltonian by taking, for instance, $h_i$ to be positive infinity, so that
\be
H_P\to h_i  \mathbb{1} \otimes \sigma_i\,,~~\text{no summation over $i$}  \,.
\ee	
Considering bilayer models for simplicity. Then we have eigen equiation
\be
\left(\begin{array}{cc}h_i\sigma_i-E_{\pm} & 0 \\0 & h_i\sigma_i-E_{\pm}\end{array}\right)\left(\begin{array}{c}\xi_{\pm} \\ \eta_{\pm}\end{array}\right)=0\,,
\ee
with solutions
\be
&&E_{\pm}=\pm h_i,\nonumber\\
&&\xi_{\pm}\xi_{\pm}^{\dagger}=|c_1|^2 \frac{(\mathbb{1} \pm \sigma_i)}{2},\nonumber\\
&&\eta_{\pm}\eta_{\pm}^{\dagger}=|c_2|^2 \frac{(\mathbb{1} \pm \sigma_i)}{2},\nonumber\\
&&|c_1|^2+|c_2|^2=1.
\ee
So a pure initial state has density matrix as
\be \label{pureini}
\rho^p_{\pm}&=&\left(\begin{array}{c}\xi_{\pm} \\ \eta_{\pm}\end{array}\right)\left(\begin{array}{c}\xi_{\pm} \\ \eta_{\pm}\end{array}\right)^{\dagger}\nonumber\\
&=& \left(\begin{array}{c}c_1 \\ c_2\end{array}\right)\left(\begin{array}{cc}c_1^* & c_2^*\end{array}\right)\otimes \frac{(\mathbb{1} \pm \sigma_i)}{2}\,,
\ee
while a mixed density matrix containing half $\left(\begin{array}{c}\xi_{\pm} \\ 0\end{array}\right)$ and half $\left(\begin{array}{c}0 \\ \eta_{\pm}\end{array}\right)$ is 
\be \label{mixini}
\rho^m_{\pm}&=&\frac{1}{2}\left(\begin{array}{c}\xi_{\pm} \\ 0\end{array}\right)\left(\begin{array}{c}\xi_{\pm} \\ 0\end{array}\right)^{\dagger}+\frac{1}{2}\left(\begin{array}{c}0 \\ \eta_{\pm}\end{array}\right)\left(\begin{array}{c}0 \\ \eta_{\pm}\end{array}\right)^{\dagger}\nonumber\\
&=& \frac{\mathbb{1}}{2}\otimes \frac{(\mathbb{1} \pm \sigma_i)}{2}.
\ee
\subsection{Derivation of The Common BIS}
\label{lem}
The TASP for any Hamiltonian can be written as 
\be\label{tast}
\overline{\langle  \mathbb{1} \otimes \sigma_i \rangle}_{\rho_0}&=&\lim_{T\to \infty}
\frac{1}{T} \int_0^T ~dt ~\text{Tr} [ \rho_0 e^{iHt} \mathbb{1} \otimes \sigma_i e^{-iHt} ] \nonumber\\
&=& \sum_m \langle \widetilde{\psi}_m | \rho_0 | \widetilde{\psi}_m\rangle \langle \widetilde{\psi}_m |  \mathbb{1} \otimes \sigma_i | \widetilde{\psi}_m\rangle\,.
\ee 
If a Hamiltonian has the following property
\be
&&H=H_0+h_3 \mathbb{1} \otimes \sigma_3 \,, \\
&&\{H_0,h_3 \mathbb{1} \otimes \sigma_3\}=0,
\ee
then
\be\label{idsig3}
\langle \widetilde{\psi}_m |  \mathbb{1} \otimes \sigma_3 | \widetilde{\psi}_m\rangle=h_3/E_m.
\ee
where $ | \widetilde{\psi}_m\rangle$ is the normalized eigenvector, $E_m$ is the eigenvalue of $H$ and index $m$ represent indeces $r$ and $\pm$.
This is because on one hand
\begin{footnotesize}
	\be
	\langle \widetilde{\psi}_m | \{H, \mathbb{1} \otimes \sigma_3\} | \widetilde{\psi}_m\rangle=\langle \widetilde{\psi}_m | \{h_3 \mathbb{1} \otimes \sigma_3, \mathbb{1} \otimes \sigma_3\} | \widetilde{\psi}_m\rangle=2h_3,\nonumber\\
	\ee
\end{footnotesize}
on the other hand 
\be
\langle \widetilde{\psi}_m | \{H, \mathbb{1} \otimes \sigma_3\} | \widetilde{\psi}_m\rangle=2E_m \langle \widetilde{\psi}_m |  \mathbb{1} \otimes \sigma_3 | \widetilde{\psi}_m\rangle.
\ee
Substituting Eq. (\ref{idsig3}) and (\ref{mixini}) into Eq. (\ref{tast}) we get
\be
\overline{\langle  \mathbb{1} \otimes \sigma_i \rangle}_{\rho_0}=-h_3 \sum_m \frac{\langle \widetilde{\psi}_m |  \mathbb{1} \otimes \sigma_i | \widetilde{\psi}_r\rangle}{4E_m},
\ee 
Note that if we choose a pure state like Eq. (\ref{pureini}), in general, we will not get a form like
\be
\overline{\langle  \mathbb{1} \otimes \sigma_i \rangle}_{\rho_0}\propto-h_3.
\ee
so it is important that the initial state is chosen as Eq. (\ref{mixini}).
\subsection{Calculation of time-averaged spin polarization}
\label{timeaverage}
\subsubsection{AB\&BA systems}
First we calculate the TASP in the bilayer case. Using Eq. (\ref{idsig3}) and (\ref{tast}), we calculate
\be
&&\overline{\langle  \mathbb{1} \otimes \sigma_i \rangle}_{\rho_0}\nonumber\\&=&\lim_{T\to \infty}
\frac{1}{T} \int_0^T ~dt ~\text{Tr} [ \rho_0 e^{iH^2_{AB\&BA}t} \mathbb{1} \otimes \sigma_i e^{-iH^2_{AB\&BA}t} ] \nonumber\\
&=&\lim_{T\to \infty}
\frac{1}{T} \int_0^T ~dt ~\text{Tr} [ S\rho_0S^{-1} e^{iSH^2_{AB\&BA}S^{-1}t}  \nonumber\\
&&S\mathbb{1} \otimes \sigma_i S^{-1}e^{-iSH^2_{AB\&BA}S^{-1}t} ] \nonumber\\
&=&\lim_{T\to \infty}
\frac{1}{T} \int_0^T ~dt ~\text{Tr} [ \rho_0 e^{i{H^2_{AB\&BA}}'t} \mathbb{1} \otimes \sigma_i e^{-i{H^2_{AB\&BA}}'t} ]\label{idotsiH'}\\
&=&\lim_{T\to \infty}
\frac{1}{T} \int_0^T ~dt ~\text{Tr} [ \rho^I_0 e^{iH_It}  \sigma_i e^{-iH_It} +\rho^I_0 e^{iH_{II}t}  \sigma_i e^{-iH_{II}t}]\nonumber\\
&=&-\frac{h^i_Ih^j_I}{2E^2_I}-\frac{h^{i}_{II}h^{j}_{II}}{2E^2_{II}}, \label{TAsiABBA}
\ee 
where we use

\be
S&=&\frac{\mathbb{1}+i\sigma_2}{\sqrt{2}}\otimes \mathbb{1},\nonumber\\
S \rho_0 S^{-1}&=& \frac{\mathbb{1}+i\sigma_2}{\sqrt{2}}\otimes \mathbb{1}  \Big[\frac{\mathbb{1}}{2}\otimes \frac{1}{2}(\mathbb{1}-\sigma_j)\Big] \frac{\mathbb{1}-i\sigma_2}{\sqrt{2}}\otimes\mathbb{1}\nonumber=\rho_0,\\
\rho_0&=&\frac{\mathbb{1}-\sigma_j}{4}\oplus\frac{\mathbb{1}-\sigma_j}{4}=\rho^I_0\oplus\rho^{II}_0,\\
h^1_I&=&h_1+t\,,h^1_{II}=h_1-t,\\
h^2_I&=&h_2\,,h^3_I=h_3\,,h^2_{II}=h_2,h^3_{II}=h_3.
\ee
$H^2_{AB\&BA}$ can be block-diagonalized by the transformation $S$, so we will take advantage of this to organize the quench process as well.
In order to set them apart, we
consider the operators $\mathcal{O}^i_{I}$ and $\mathcal{O}^i_{II}$ to make 
\be
\overline{\langle  \mathcal{O}^i_I \rangle}_{\rho_0}&=&-\frac{h^i_Ih^j_I}{2E^2_I},\\
\overline{\langle  \mathcal{O}^i_{II} \rangle}_{\rho_0}&=&-\frac{h^{i}_{II}h^{j}_{II}}{2E^2_{II}}.
\ee
This can be done by considering the two subspaces of $\mathbb{1} \otimes \sigma_i $ in Eq. (\ref{idotsiH'}): $\left(\begin{array}{cc}1 &  \\ & 0\end{array}\right) \otimes \sigma_i $ and  $\left(\begin{array}{cc}0 &  \\ & 1\end{array}\right) \otimes \sigma_i $, such that
\be
\label{O1ids}S \mathcal{O}^i_{I} S^{-1}&=& \left(\begin{array}{cc}2 &  \\ & 0\end{array}\right) \otimes \sigma_i,\\
\label{O2ids}S \mathcal{O}^i_{II} S^{-1}&=& \left(\begin{array}{cc}0 &  \\ & 2\end{array}\right) \otimes \sigma_i,
\ee
The above two equations can be solved by: 
\be
\mathcal{O}^i_I=(\mathbb{1}+\sigma_1)\otimes \sigma_i,\\
\mathcal{O}^i_{II}=(\mathbb{1}-\sigma_1)\otimes \sigma_i. 
\ee

\label{MulayABBAapp}
Second, we calculate the TASP in the multilayer case.
\be
&&\overline{\langle  \mathbb{1} \otimes \sigma_i \rangle}_{\rho_0}\nonumber\\
&=&\lim_{T\to \infty}
\frac{1}{T} \int_0^T ~dt ~\text{Tr} [ \rho_0 e^{iH^n_{AB\&BA}t} \mathbb{1} \otimes \sigma_i e^{-iH^n_{AB\&BA}t} ] \nonumber\\
&=&\lim_{T\to \infty}
\frac{1}{T} \int_0^T ~dt ~\text{Tr} [ S\rho_0S^{-1} e^{iSH^n_{AB\&BA}S^{-1}t} \nonumber\\
&&S\mathbb{1} \otimes \sigma_i S^{-1}e^{-iSH^n_{AB\&BA}S^{-1}t} ] \nonumber\\
&=&\lim_{T\to \infty}
\frac{1}{T} \int_0^T ~dt ~\text{Tr} [ \rho_0 e^{i{H^n_{AB\&BA}}'t} \mathbb{1} \otimes \sigma_i e^{-i{H^n_{AB\&BA}}'t} ]\label{idotsinH'}\nonumber\\
\\
&=&\lim_{T\to \infty}
\frac{1}{T} \int_0^T ~dt ~\text{Tr} [ \sum_{r=1}^n\rho^I_0 e^{iH_rt}  \sigma_i e^{-iH_rt} ]\nonumber\\
&=&-\sum_{r=1}^n\frac{h^i_rh^j_r}{2E^2_r}, \label{TAsinABBAapp}
\ee 
where we use
\be
S\rho_0S^{-1}&=&\rho_0, S=S_{\Sigma_1}\otimes \mathbb{1},\\
\rho_0&=&\oplus_{r=1}^N \rho^r_0,   \rho^r_0=\frac{\mathbb{1}-\sigma_j}{2N},\\
h^1_r&=&h_1-2t\cos\theta_r,\\
h^2_r&=&h_2\,,\,h^3_r=h_3.
\ee
In order to set them apart, we
consider  the operators $\mathcal{O}_r$  such that 
\be
\label{O1hr}\overline{\langle  \mathcal{O}^i_r \rangle}_{\rho_0}&=&-\frac{h^i_rh^j_r}{E^2_r}.
\ee
This can be done by considering the N subspaces of $\mathbb{1} \otimes \sigma_i $ in Eq. (\ref{idotsiH'}): such that
\be
\label{Orids}S \mathcal{O}^i_r S^{-1}&=& (\underbrace{\mathbf{0}\oplus...\oplus\mathbf{0}}_{r-1~ \mathbf{0}s}\oplus \mathbb{1}_2 \oplus \mathbf{0}\oplus...\mathbf{0}) \otimes \sigma_i,
\ee
which is solved by
\be
\mathcal{O}^i_r=\mathbf{a}_r\mathbf{a}_r^T\otimes \sigma_i. 
\ee
Here, $\mathbf{a}_r$ is defined in Eq. (\ref{eq:ar}).

\subsubsection{BA systems}
\label{BAbiapp}

From Eq. (\ref{tast}), Eq. (\ref{tpidsig1tp}) and (\ref{tpidsig2tp}), for bilayer system, the components of TASP are as following:
\be
\overline{\langle  \mathbb{1} \otimes \sigma_1 \rangle}_{\rho_0}
&=&-h_1h_3\sum_m \frac{E_m^2-h_3^2}{2E_m^2(E_m^2-h_3^2+h_1^2+h_2^2)}\nonumber\\
\overline{\langle  \mathbb{1} \otimes \sigma_2 \rangle}_{\rho_0}
&=&-h_2h_3\sum_m \frac{E_m^2-h_3^2}{2E_m^2(E_m^2-h_3^2+h_1^2+h_2^2)}\\
\overline{\langle  \mathbb{1} \otimes \sigma_3 \rangle}_{\rho_0}&=&-h_3^2 \sum_m \frac{1}{4E_m^2}\nonumber.
\ee
in which the initial state $$\rho_0=\frac{\mathbb{1}}{2}\otimes \frac{1}{2}(\mathbb{1}-\sigma_3).$$
For the initial state $$\rho_0=\frac{\mathbb{1}}{2}\otimes \frac{1}{2}(\mathbb{1}-\sigma_a)\,, a=1,2.$$ the components of TASP are :
\be
\overline{\langle  \mathbb{1} \otimes \sigma_1 \rangle}_{\rho_0}
&=&-h_1h_a\sum_m \frac{(E_m^2-h_3^2)^2}{E_m^2(E_m^2-h_3^2+h_1^2+h_2^2)^2},\nonumber\\
\overline{\langle  \mathbb{1} \otimes \sigma_2 \rangle}_{\rho_0}
&=&-h_2h_a\sum_m \frac{(E_m^2-h_3^2)^2}{E_m^2(E_m^2-h_3^2+h_1^2+h_2^2)^2}\\
\overline{\langle  \mathbb{1} \otimes \sigma_3 \rangle}_{\rho_0}&=&-h_3h_a \sum_m \frac{E_m^2-h_3^2}{2E_m^2(E_m^2-h_3^2+h_1^2+h_2^2)}\nonumber.	
\ee
In short, we have 
\be
\overline{\langle  \mathbb{1} \otimes \sigma_i \rangle}_{\rho_0}&=&-h_ih_jA_{ij}~~\text{(no summation over i,j)},
\\
\rho_0&=&\frac{\mathbb{1}}{2}\otimes \frac{1}{2}(\mathbb{1}-\sigma_j),\\
\label{A12}A_{11}=A_{12}&=&A_{21}=A_{22}=\sum_m \frac{(E_m^2-h_3^2)^2}{E_m^2(E_m^2-h_3^2+h_1^2+h_2^2)^2},\nonumber\\ \\
\label{A13}A_{13}=A_{31}&=&A_{23}=A_{32}=\sum_m \frac{E_m^2-h_3^2}{2E_m^2(E_m^2-h_3^2+h_1^2+h_2^2)},\nonumber\\ \\
\label{A33}A_{33}&=&\sum_m \frac{1}{4E_m^2}\,.
\ee
Let us look at the dynamical spin-texture fields
\be
\widetilde{g_i(\mathbf{k})}=-\frac{1}{\mathcal{N}_\mathbf{k}}\partial_{k_{\bot}}\overline{\langle  \mathbb{1} \otimes \sigma_i \rangle}_{\rho_0}
\ee
For initial state $$\rho_0=\frac{\mathbb{1}}{2}\otimes \frac{1}{2}(\mathbb{1}-\sigma_3)\,,$$ whose BIS is at $h_3=0$, the difference is calculated as
\be
&&\Delta \overline{\langle  \mathbb{1} \otimes \sigma_a \rangle}_{\rho_0}\Big|_{k_{\bot}\to0}\nonumber\\
&=&-\Big((h_ah_3\sum_m \frac{E_m^2-h_3^2}{2E_m^2(E_m^2-h_3^2+h_1^2+h_2^2)})\Big|_{(k_{\bot},\mathbf{k}_{\parallel})}\nonumber\\
&-&(h_ah_3\sum_m \frac{E_m^2-h_3^2}{2E_m^2(E_m^2-h_3^2+h_1^2+h_2^2)})\Big|_{(-k_{\bot},\mathbf{k}_{\parallel})} \Big)\Big|_{k_{\bot}\to0}\nonumber\\
&\propto&-2h_ak_\bot \sum_m\frac{E_m^2-h_3^2}{2E_m^2(E_m^2-h_3^2+h_1^2+h_2^2)})\Big|_{(0,\mathbf{k}_{\parallel})}.
\ee
$a=1,2$.
After normalization, we get
\be
\widetilde{g_{a}(\mathbf{k})}\Big|_{\mathbf{k}\in \text{BIS}}=\frac{h^a(0,\mathbf{k}_\parallel)}{\sum_{a=1}^2 (h^a(0,\mathbf{k}_\parallel))^2}=\hat{h}^{so,a}, 
\ee
For initial state $$\rho_0=\frac{\mathbb{1}}{2}\otimes \frac{1}{2}(\mathbb{1}-\sigma_a)\,,a=1,2.$$ whose BIS is at $h_a=0$,  the difference is calculated as
\begin{scriptsize}
	\be
	\Delta \overline{\langle  \mathbb{1} \otimes \sigma_b \rangle}_{\rho_0}\Big|_{k_{\bot}\to0}	
	&\propto&-2h_bk_\bot \sum_m \frac{(E_m^2-h_3^2)^2}{E_m^2(E_m^2-h_3^2+h_1^2+h_2^2)^2}\Big|_{(0,\mathbf{k}_{\parallel})}, b=1,2.\nonumber
	\ee
\end{scriptsize}
.
\be
\Delta \overline{\langle  \mathbb{1} \otimes \sigma_3 \rangle}_{\rho_0}\Big|_{k_{\bot}\to0}	
&\propto&-2h_3k_\bot \sum_m \frac{E_m^2-h_3^2}{2E_m^2(E_m^2-h_3^2+h_1^2+h_2^2)}\Big|_{(0,\mathbf{k}_{\parallel})}.\nonumber\\
\ee
In short, for $$\rho_0=\frac{\mathbb{1}}{2}\otimes \frac{1}{2}(\mathbb{1}-\sigma_j)\,,j=1,2,3.$$ 
\be
\Delta \overline{\langle  \mathbb{1} \otimes \sigma_i \rangle}_{\rho_0}\Big|_{k_{\bot}\to0}
&\propto&-2(A_{ij}h_i)\Big|_{(0,\mathbf{k}_{\parallel})}k_\bot,
\ee
Thus 
\be
\widetilde{g_{p}(\mathbf{k})}\Big|_{\mathbf{k}\in \text{BIS}}&=&\frac{A_{pj}h_p}{(A_{pj}h_p)^2+(A_{qj}h_q)^2},\, p\ne q\ne j; p,q,j=1,2,3.
\ee
Here, $\widetilde{\bm {g}}$ is a vecor with two components.

Next we consider N-layer BA systems. The components of TASP are as following:
\be
\overline{\langle  \mathbb{1} \otimes \sigma_1 \rangle}_{\rho_0}&=&-h_1h_3\sum_m \frac{E_m^2-h_3^2}{ND_m},\nonumber\\
\overline{\langle  \mathbb{1} \otimes \sigma_2 \rangle}_{\rho_0}&=&-h_2h_3\sum_m \frac{E_m^2-h_3^2}{ND_m},\\
\overline{\langle  \mathbb{1} \otimes \sigma_3 \rangle}_{\rho_0}&=&-h_3^2 \sum_m \frac{1}{2NE_m^2}\nonumber.
\ee
with the initial state $$\rho_0=\frac{\mathbb{1}}{N}\otimes \frac{1}{2}(\mathbb{1}-\sigma_3).$$
For the initial state $$\rho_0=\frac{\mathbb{1}}{N}\otimes \frac{1}{2}(\mathbb{1}-\sigma_a)\,, a=1,2.$$ the components of TASP are :
\be
\overline{\langle  \mathbb{1} \otimes \sigma_1 \rangle}_{\rho_0}
&=&-h_1h_a\sum_m \frac{2(E_m^2-h_3^2)^2}{N^2D_m^2},\nonumber
\\
\overline{\langle  \mathbb{1} \otimes \sigma_2 \rangle}_{\rho_0}
&=&-h_2h_a\sum_m \frac{2(E_m^2-h_3^2)^2}{N^2D_m^2},\\
\overline{\langle  \mathbb{1} \otimes \sigma_3 \rangle}_{\rho_0}&=&-h_3h_a \sum_m \frac{E_m^2-h_3^2}{ND_m}.\nonumber
\ee
where
\be
D_m&=&[E_m(E_m(E_m^2-h_3^2+h_1^2+h_2^2)\nonumber\\
&-&\frac{1+(-1)^{N+1}}{N+1}h_3(E_m^2-h_3^2-h_1^2-h_2^2)).
\ee
In short, we have 
\be
\overline{\langle  \mathbb{1} \otimes \sigma_i \rangle}_{\rho_0}&=&-h_ih_jA^{(N)}_{ij}~~\text{(no summation over i,j)},\nonumber\\
\\
\rho_0&=&\frac{\mathbb{1}}{2}\otimes \frac{1}{2}(\mathbb{1}-\sigma_j)\\
\label{An12}A^{(N)}_{11}&=&A^{(N)}_{12}=A^{(N)}_{21}=A^{(N)}_{22},\nonumber\\
&=&\sum_m \frac{2(E_m^2-h_3^2)^2}{D_m^2},\nonumber\\
\\
\label{An13}A^{(N)}_{13}&=&A^{(N)}_{31}=A^{(N)}_{23}=A^{(N)}_{32}\nonumber\\
&=&\sum_m \frac{E_m^2-h_3^2}{D_m},\nonumber\\
\\
\label{An33}A^{(N)}_{33}&=&\sum_m \frac{1}{2NE_m^2}.
\ee
Like bilayer case, for the initial state $$\rho_0=\frac{\mathbb{1}}{2}\otimes \frac{1}{2}(\mathbb{1}-\sigma_j)\,,j=1,2,3.$$ 
\be
\Delta \overline{\langle  \mathbb{1} \otimes \sigma_i \rangle}_{\rho_0}\Big|_{k_{\bot}\to0}
&\propto&-2(A^{(N)}_{ij}h_i)\Big|_{(0,\mathbf{k}_{\parallel})}k_\bot.
\ee
and 
\be
\widetilde{g_{p}(\mathbf{k})}\Big|_{\mathbf{k}\in \text{BIS}}&=&\frac{A^{(N)}_{pj}h_p}{(A^{(N)}_{pj}h_p)^2+(A^{(N)}_{qj}h_q)^2},\, p\ne q\ne j; p,q,j=1,2,3.\nonumber 
\ee
Here, $\widetilde{\bm {g}}$ is a vecor with two components.

\end{document}